\documentclass[12pt]{article}
\usepackage{epsfig}
\usepackage{amssymb}
\newcommand{\bfig}{\begin{center}\begin{picture}}
\newcommand{\efig}[1]{\end{picture}\\{\small #1}\end{center}}

\newcommand{\bmip}[2]{\begin{minipage}[t]{#1pt}\bfig(#1,#2)}
\newcommand{\emip}[1]{\efig{#1}\end{minipage}}

\newcommand{\bq}{\begin{equation}}
\newcommand{\eq}{\end{equation}}
\newcommand{\bqa}{\begin{eqnarray}}
\newcommand{\eqa}{\end{eqnarray}}

\newcommand{\eqn}[1]{eq.~(\ref{#1})}

\newcommand{\gev}{\mbox{GeV}}
\newcommand{\mev}{\mbox{MeV}}
        \newdimen\eqskip
        \newdimen\txtskip
        \baselineskip=\txtskip
        \newdimen\mysep                
        \newdimen\hmysep
\begin{document}
       
  \newcommand{\ccaption}[2]{
    \begin{center}
    \parbox{0.85\textwidth}{
      \caption[#1]{\small{{#2}}}
      }
    \end{center}
    }
\newcommand{\BS}{\bigskip}
\def    \be             {\begin{equation}}
\def    \ee             {\end{equation}}
\def    \beq             {\begin{equation}}
\def    \eeq             {\end{equation}}
\def    \ba             {\begin{eqnarray}}
\def    \ea             {\end{eqnarray}}
\def    \beqn           {\begin{eqnarray}}
\def    \eeqn           {\end{eqnarray}}
\def    \beeq           {\begin{eqnarray}}
\def    \eeeq           {\end{eqnarray}}
\def    \nn             {\nonumber}
\def    \=              {\;=\;}
\def    \frac           #1#2{{#1 \over #2}}
\def    \ret            {\\[\eqskip]}
\def    \ie             {{\em i.e.\/} }
\def    \eg             {{\em e.g.\/} }
\def \lsim{\mathrel{\vcenter
     {\hbox{$<$}\nointerlineskip\hbox{$\sim$}}}}
\def \gtrsim{\mathrel{\vcenter
     {\hbox{$>$}\nointerlineskip\hbox{$\sim$}}}}
\def    \bentarrow      {\:\raisebox{1.1ex}{\rlap{$\vert$}}\!\rightarrow}
\def    \rd             {{\mathrm d}}    
\def    \Im             {{\mathrm{Im}}}  
\def    \bra#1          {\mbox{$\langle #1 |$}}
\def    \ket#1          {\mbox{$| #1 \rangle$}}

\def    \kev            {\mbox{$\mathrm{keV}$}}
\def    \mev            {\mbox{$\mathrm{MeV}$}}
\def    \gev            {\mbox{$\mathrm{GeV}$}}


\def    \mq             {\mbox{$m_Q$}}  
\def    \mt             {\mbox{$m_t$}}  
\def    \mb             {\mbox{$m_b$}}  
\def    \mqq            {\mbox{$m_{Q\bar Q}$}}
\def    \mqqsq          {\mbox{$m^2_{Q\bar Q}$}}
\def    \pt             {\mbox{$p_T$}}
\def    \et             {\mbox{$E_T$}}
\def    \xt             {\mbox{$x_T$}}
\def    \xtsq           {\mbox{$x_T^2$}}
\def    \ptsq           {\mbox{$p^2_T$}}
\def    \etsq           {\mbox{$E^2_T$}}
\def    \rjet           {\mbox{$R_{\mathrm jet}$}}
\newcommand     \MSB            {\ifmmode {\overline{\rm MS}} \else 
                                 $\overline{\rm MS}$  \fi}
\def    \muf            {\mbox{$\mu_{\rm F}$}}
\def    \mug            {\mbox{$\mu_\gamma$}}
\def    \mufsq          {\mbox{$\mu^2_{\rm F}$}}
\def    \mur            {{\mbox{$\mu_{\rm R}$}}}
\def    \mursq          {\mbox{$\mu^2_{\rm R}$}}
\def    \mul            {{\mu_\Lambda}}
\def    \mulsq          {\mbox{$\mu^2_\Lambda$}}

\def    \bzero          {\mbox{$b_0$}}
\def    \as             {\ifmmode \alpha_s \else $\alpha_s$ \fi}
\def    \asb            {\mbox{$\alpha_s^{(b)}$}}
\def    \assq           {\mbox{$\alpha_s^2$}}
\def \oacube {\mbox{$ {\cal O}(\alpha_s^3)$}}
\def \oaemacube {\mbox{$ {\cal O}(\alpha\alpha_s^3)$}}
\def \oafour {\mbox{$ {\cal O} (\alpha_s^4)$}}
\def \oatwo {\mbox{$ {\cal O} (\alpha_s^2)$}}
\def \oaematwo {\mbox{$ {\cal O}(\alpha \alpha_s^2)$}}
\def \oaemas {\mbox{$ {\cal O}(\alpha \alpha_s)$}} 
\def \oas   {\mbox{$ {\cal O}(\alpha_s)$}}
\def\slash#1{{#1\!\!\!/}}
\def\rt1{\raisebox{-1ex}{\rlap{$\; \rho \to 1 \;\;$}}
\raisebox{.4ex}{$\;\; \;\;\simeq \;\;\;\;$}}
\def\ltap{\raisebox{-.5ex}{\rlap{$\,\sim\,$}} \raisebox{.5ex}{$\,<\,$}}
\def\gtap{\raisebox{-.5ex}{\rlap{$\,\sim\,$}} \raisebox{.5ex}{$\,>\,$}} 

\newcommand\LambdaQCD{\Lambda_{\scriptscriptstyle \rm QCD}}

\def\onebj{1-$b$-jet}
\def\bjet{$b$ jet}
\def\twobj{2-$b$-jet}
\def\ppbar{\ifmmode p{\buildrel {
{\scriptscriptstyle(}-{\scriptscriptstyle )}}  \over {p}} \else 
           $p {\buildrel {
{\scriptscriptstyle(}-{\scriptscriptstyle )}}  \over {p}}$ \fi}
\def\naive{na\"{\i}ve}
\def\asp{{\alpha_s}\over{\pi}}
\def\half{\frac{1}{2}}
\def\herwig{\small HERWIG}
\def\isajet{\small ISAJET}
\def\pythia{\small PYTHIA}
\def\grace{\small GRACE}
\def\vecbos{\small VECBOS}
\def\madgraph{\small MADGRAPH}
\def\comphep{\small CompHEP}
\def\ALPHA{\small ALPHA}
\def\met{$\rlap{\kern.2em/}E_T$}
\begin{titlepage}
\nopagebreak
{\flushright{
        \begin{minipage}{5cm}
        CERN-TH/2001-215\\
        {\tt hep-ph/0108069}
        \end{minipage}        }
        
}
\vfill
\begin{center}
{ { \bf \sc \Huge
Multijet matrix
elements \\ and shower evolution \\ in hadronic collisions: \\[2mm]
$Wb\bar{b}+n$ jets as a case study.}
\footnote{Presented at the FNAL Workshop on Monte-Carlos for Run~2,
  April 2001, Batavia, IL.}$^{,}$
\footnote{This work
was supported in part  by the EU Fourth Framework Programme ``Training and
Mobility  of Researchers'',  Network ``Quantum Chromodynamics and the Deep
Structure of Elementary Particles'', contract FMRX--CT98--0194 (DG 12 --
MIHT).}

}
\vfill                                                       
{\bf          Michelangelo L. MANGANO$^{(a)}$
          Mauro MORETTI$^{(b)}$ and
          Roberto PITTAU$^{(c)}$  }
\\[0.5cm]
{\small
$^{(a)}$ CERN, Theoretical Physics Division, CH~1211 Geneva 23, Switzerland
\\
$^{(b)}$ Dipartimento di Fisica, Universit\`{a} di Ferrara, and INFN, 
         Ferrara, Italy
\\
$^{(c)}$ Dipartimento di Fisica, Universit\`{a} di Torino, and INFN, 
         Turin, Italy
}
\end{center}                                   
\nopagebreak
\vfill
\begin{abstract} 
We study in this paper the  production, in hadronic
collisions, of final states with $W$ gauge bosons, 
 heavy quark pairs and $n$ extra jets (with $n$ up to 4). The complete
  partonic tree-level QCD matrix elements are evaluated using the \ALPHA\
 algorithm, and the events generated at the parton level are then
 evolved through the QCD shower and  eventually hadronised using the
 coherent shower evolution provided by the \herwig\ Monte Carlo. We
 discuss the details of our Monte Carlo implementation, and present
 results of phenomenological interest for the 
 Tevatron Collider and for the LHC. We also comment on the impact of
 our calculation on the backgrounds to $W (H\to b\bar{b})$
 production, when only one $b$ jet is tagged.
\end{abstract}                                                
\vskip 1cm
CERN-TH/2001-215\hfill \\
\today \hfill  
\vfill       
\end{titlepage}

\section{Introduction}
\label{sec:intro}
Multijet final states are characteristic of a large class of important
phenomena present in high-energy collisions. QCD interactions generate
multijet final states via radiative processes 
at high orders of perturbation theory. Heavy
particles in the Standard Model (SM), 
such as $W$ and $Z$ bosons or the top quark, $t$, decay to 
multiquark configurations (eventually leading to jets) 
via electroweak (EW) interactions. In addition to the above SM
sources, particles possibly present in theories beyond the SM
(BSM) are expected to decay to multiparton final states, and therefore
to lead to multijets. Typical examples are the cascade decays to
quarks and gluons of supersymmetric strongly interacting particles,
such as squarks and gluinos. 

Whether our interest is in accurate measurements of top quarks or in
the search for more exotic states, multijet final states always
provide an important observable, and the study of the backgrounds due
to QCD is an essential part of any experimental analysis.

In addition to fully hadronic multijet final states, a special
interest exists in final states where the jets are accompanied by gauge
bosons. For example, the associated production of several jets with
a $Z$ boson decaying to neutrinos gives rise to multijet+missing
transverse energy (\met) final states. These provide an important
background to the search for supersymmetric particles. Likewise, the
associated production of 4 jets and a $W$ boson decaying leptonically
provides the leading source of backgrounds to the identification and
study of top quark pairs in hadronic collisions.

Several examples of calculation of multijet cross-sections in hadronic
collisions exist in the literature. Some of them are included in
parton-level Monte Carlo (MC) event generators, where final states
consisting of hard and well isolated partons are generated. Among the
most used and best documented examples are
PAPAGENO~\cite{Hinchliffe:1993de} (a compilation of several partonic
processes), VECBOS~\cite{Berends:1991ax} (for production of $W/Z$
bosons in association with up to 4 jets), NJETS~\cite{Berends:1989ie}
(for production of up to 6 jets). In addition to these ready-to-use
codes, programs for the automatic generation of user-specified
parton-level processes exist, and have been used for the calculation
of many important processes in hadronic collisions:
\madgraph~\cite{Stelzer:1994ta}, \comphep~\cite{Pukhov:1999gg} and
\grace~\cite{Ishikawa:1993qr}.

Studies of the partonic final states can be performed by assuming that
each hard parton can be identified with a jet, and that the jets'
momenta are equal to those of the parent parton.  This simplification is
extremely useful to get rough estimates of production rates, but
cannot be used in the context of realistic detector simulations, for
which a representation of the full structure of the final state (in
terms of hadrons) is required.  This full description can be obtained
by merging the partonic final states with so-called shower MC
programs (such as \herwig~\cite{Marchesini:1988cf},
\pythia~\cite{Sjostrand:1994yb} or \isajet~\cite{Paige:1998xm}), 
where partons are perturbatively evolved through
emission of gluons, and subsequently hadronized.
As we shall discuss in the following, however, this merging is not always
possible, since common parton-level MC's sum and average over
flavours and colours, and do not usually provide sufficient
information on the flavour and colour content of the events.

The goal of the present work is twofold. First we review a general
strategy proposed in~\cite{Caravaglios:1999yr}
for the construction of event generators for multijet final
states, based on the exact leading-order (LO) evaluation of the matrix
elements for assigned flavour and colour configurations, 
and the subsequent shower development and transition into a
fully hadronized final state.  In view of the complexity of the
problem, we shall argue in favour of the use of the algorithm 
\ALPHA~\cite{Caravaglios:1995cd} 
as the best possible tool to carry out the relative 
matrix-element computations.
Applications of \ALPHA\ to the case of hadronic collisions have
already been shown successful in the case of multijet production
in~\cite{Draggiotis:1998gr,Caravaglios:1999yr}.

Then we apply the technique to the specific example of production of
$W Q\bar{Q}+n$-jet final states (with $Q$ being a massive quark, and
$n \le 4$). The matrix-element calculation and the generation of
parton-level events is completed with the shower evolution and
hadronization generated by the \herwig\ MC~\cite{Marchesini:1988cf}.
We present the results, in the case of $Q=b$, of the parton-level
calculations for several production rates and distributions of
interest at the Tevatron and at the LHC.  We discuss some interesting
features of the final states, exploring in particular the relative
size of processes where the $b$ and $\bar{b}$ give rise to either 1 or
2 taggable jets.  This comparison will underscore the importance of
full matrix-element calculations taking fully into account the
heavy-quark mass effects.  We then present some results relative to
final states reconstructed after the shower evolution. We compare
parton-level to jet-level distributions, and study the ability of
\herwig\ to approximate the emission rate of extra jets via the shower
evolution.
Independent work on the merging of parton-level calculations with
shower MC's (for the specific case of \pythia) has been pursued by the
\comphep\ group~\cite{Belyaev:2000wn} and by the \grace\
group~\cite{Sato:2001ae}, 
limited however to final states with at most 4 partons.

We conclude this presentation by listing future projects which could
be realized within the framework of the approach presented here.

\section{Matrix-element evaluation}
The emission of soft gluon radiation in state-of-the-art
shower-evolution programs accounts for quantum coherence, which is
implemented via the prescription of angular ordering in the parton
cascade~\cite{Bassetto:1983ik}.  Angular ordering dictates that the
radiation emitted by a colour dipole be confined within the cone
defined by the directions of the two colour charges defining the
dipole itself.  The set of colour connections among the partons which
defines the set of dipoles for a given event will be called a {\it
  colour flow}, or {\it colour configuration}.

The comparison with existing hadron-collider data~\cite{Abe:1994nj},
confirms that the constraint of angular ordering is essential to
properly describe the particle multiplicity and the momentum
distribution inside jets, as well as to describe the correlations
between primary jets and softer jets emitted during the shower
evolution. In order to reliably evolve a multiparton state into a
multijet configuration, it is therefore necessary to associate a
specific colour-flow pattern to each generated parton-level event.
This requires an ad-hoc approach to the evaluation of the matrix
elements. A specific proposal was presented
in~\cite{Caravaglios:1999yr}, and will be shortly summarized here.

\subsection{Reconstruction of colour flows}
We discuss for simplicity the case of multigluon
processes~\cite{Mangano:1988xk}, as the extensions to cases with
quarks and electroweak particles~\cite{Mangano:1988kp,Mangano:1991by}
follow the same pattern.  The scattering amplitude for $n$ gluons with
momenta $p_i^{\mu}$, helicities $\epsilon_i^{\mu}$ and colours $a_i$
(with $i=1,\dots,n$), can be written as~\cite{Mangano:1988xk}:
\be      \label{eq:dualg}
  M(\{p_i\},\{\epsilon_i\},\{a_i\}) \; = \;
  \sum_{P(2,3,\dots,n)} \; \mbox{tr}(\lambda^{a_{i_1}}\,\lambda^{a_{i_2}}\dots
  \lambda^{a_{i_n}}) \; A(\{p_{i_1}\},\{\epsilon_{i_1}\}; \dots 
  \{p_{i_n}\},\{\epsilon_{i_n}\}) \; .
\ee                                          
The sum extends over all permutations $P_i$ of $(2,3,\dots ,n)$, and the
functions $A(\{P_i\}) $ (known as {\em dual} or {\em
colour-ordered} amplitudes) are gauge-invariant,
cyclically-symmetric functions of the gluons'
momenta and helicities. Each dual amplitude $A(\{P_i\})$
corresponds to a set of diagrams where colour flows from one gluon to
the next, according to the ordering specified by the permutation of indices.
When summing over colours the amplitude squared, different
orderings of dual amplitudes are orthogonal at the leading order in $1/N^2$:
\be      \label{eq:dualgsq}
  \sum_{{\rm col's}} \vert M(\{p_i\},\{\epsilon_i\},\{a_i\}) \vert^2 \; = \;
  N^{n-2}(N^2-1)   \sum_{P_i} 
    \left[  \vert A(\{P_i\})   \vert^2  +\frac{1}{N^2}(\mathrm{interf.})
    \right]\, .
\ee                                  
At the leading order in $1/N^2$, therefore, the square of each dual
amplitude is proportional to the relative probability of the
corresponding colour flow. Each flow defines, in a gauge invariant
way, the set of colour currents which are necessary and sufficient
to implement the colour ordering prescription necessary for the
coherent evolution of the gluon shower. Because of the incoherence of
different colour flows, each event can be assigned a specific colour
configuration by comparing the relative size of $ \vert A(\{P_i\})   \vert^2 $
for all possible flows. 

When working at the physical value of $N_c=3$, the interferences among
different flows cannot be neglected in the evaluation of the square of
the matrix element, \eqn{eq:dualg}. 
As a result, the basis of colour flows does not
provide an orthogonal set of colour states, and a MC selection of
colour flows is not possible. In~\cite{Caravaglios:1999yr} we proposed
to use the Gell-Mann basis of $SU(3)$ matrices
as an orthogonal basis to represent the
colour state of a given set of partons:
\ba                                          
&&   \lambda^1 \= \frac{1}{\sqrt{2}} \left( \begin{array}{ccc}
                          0 & 1 & 0 \\
                          0 & 0 & 0 \\
                          0 & 0 & 0 \end{array} \right) \quad,\quad
     \lambda^2 \= \frac{1}{\sqrt{2}} \left( \begin{array}{ccc}
                          0 & 0 & 1 \\
                          0 & 0 & 0 \\
                          0 & 0 & 0 \end{array} \right) \quad,\quad
     \lambda^3 \= \frac{1}{\sqrt{2}} \left( \begin{array}{ccc}
                          0 & 0 & 0 \\
                          1 & 0 & 0 \\
                          0 & 0 & 0 \end{array} \right) \nn \\
&&   \lambda^5 \= \frac{1}{\sqrt{2}} \left( \begin{array}{ccc}
                          0 & 0 & 0 \\
                          0 & 0 & 1 \\
                          0 & 0 & 0 \end{array} \right) \quad,\quad
     \lambda^6 \= \frac{1}{\sqrt{2}} \left( \begin{array}{ccc}
                          0 & 0 & 0 \\
                          0 & 0 & 0 \\
                          1 & 0 & 0 \end{array} \right) \quad,\quad
     \lambda^7 \= \frac{1}{\sqrt{2}} \left( \begin{array}{ccc}
                          0 & 0 & 0 \\    
                          0 & 0 & 0 \\
                          0 & 1 & 0 \end{array} \right) \nn \\
&&   \lambda^4 \= \frac{1}{{2}} \left( \begin{array}{ccc}
                          1 & 0 & 0 \\
                          0 &-1 & 0 \\
                          0 & 0 & 0 \end{array} \right) \quad,\quad
     \lambda^8 \= \frac{1}{\sqrt{12}} \left( \begin{array}{ccc}
                          1 & 0 & 0 \\    
                          0 & 1 & 0 \\
                          0 & 0 &-2 \end{array} \right) \nn 
\ea                                                       
In this basis, only a fraction of all possible $8^n$ colour assignments gives
rise to a non-zero amplitude. 
For each event, we randomly select a non-vanishing
colour state for the external gluons,
and evaluate the amplitude $M$. We then list all dual amplitudes               
contributing to the chosen colour configuration according to
eq.~(\ref{eq:dualg}) and, among these dual 
amplitudes, we randomly select a colour flow on the basis of their
relative weight.
In a 6-gluon amplitude, for example, 
a possible non-zero colour assignment is given by              
$(2,7,5,6,1,3)$.
Up to cyclic permutations, only three orderings of the colour indices give
rise to non-vanishing traces:
${\rm  tr}(\lambda^2\,\lambda^7\,\lambda^5\,\lambda^6\,\lambda^1\,\lambda^3)$, 
${\rm  tr}(\lambda^2\,\lambda^6\,\lambda^1\,\lambda^5\,\lambda^7\,\lambda^3)$ 
and  
${\rm tr}(\lambda^2\,\lambda^7\,\lambda^3\,\lambda^1\,\lambda^5\,\lambda^6)$. 
Therefore only three dual
amplitudes contribute to the full amplitude: $A(2,7,5,6,1,3)$,
$A(2,6,1,5,7,3)$ and $A(2,7,3,1,5,6)$. The colour ordering to be specified for
the coherent parton-shower evolution can be selected by comparing the size of
the squares of ${\rm tr}(\lambda^2\,\lambda^{i_2}\dots\lambda^{i_6}) \,
A(2,i_2,\dots,i_6)$ for the three contributing permutations $(i_2,\dots,i_6)$ 
of the colour indices.            

In the limit of a large number of generated events, and in the case of
processes with only gluons or with gluons and one quark-antiquark
pair, this algorithm is equivalent to the colour-flow extraction
algorithm proposed in~\cite{Odagiri:1998ep} and employed in \herwig.
There, the full sum over all colours is performed for each event. 
For each event one then calculates the individual subamplitudes
$M(f)$ corresponding to all possible colour flows $f$.  The event is
then assigned the colour flow $\bar{f}$ with a probability:

\be P(\bar{f}) = \frac{\vert M(\bar{f})
  \vert ^2}{ \sum_{f} \vert M(f) \vert ^2} \; .
\ee 

In the case of two or more $q\bar{q}$ pairs our prescription and that
of~\cite{Odagiri:1998ep} 
differ at order $1/N_c^2$, since by generating colour states we
produce configurations whose matrix element has no leading colour contribution.
One such example is the process $q_i \bar{q}_i \to q_j
\bar{q}_j$, where $i,j=1,\dots,N_c$ are fixed colours. This transition
is mediated by the gluons corresponding to the diagonal
Gell-Mann matrices. The colour coefficient is trivially given by 
\be
\sum_{a} \; \lambda^a_{ik} \lambda^a_{jl} =
\frac{1}{2}\left(\delta_{il}\delta_{jk}-\frac{1}{N_c}\delta_{ik}\delta_{jl}
\right) \; ,
\ee
which, when $k=i$ and $l=j$ as in the proposed example, is of order $1/N_c$.
The corresponding colour flow links the quark and antiquark of the initial
state, and those of the final state. No such colour flows can appear in
the algorithm by Odagiri, where only colour configurations of leading
order in $N_c$ are generated.

Since the colour-flow information is only relevant for events which
will be evolved through a shower MC, and since one usually does this
only for unweighted events, it is sufficient to evaluate the possibly
large set of dual amplitudes corresponding to a given colour state
only for the small set of unweighted events. The extraction of a
colour flow, therefore, does not increase significantly the computing
time necessary to generate unweighted events. The full sum over colours is
obtained by averaging over a large sample of events. The cross section
thus obtained is correct to all orders in $1/N_c$. Relative to the
prescription of ~\cite{Odagiri:1998ep}, this approach has
the advantage that the random selection of colours can in principle be
optimized by exploiting the strong colour-dependence of the matrix
elements (see an example along these lines
in~\cite{Draggiotis:2000gm}). 

While the algorithm proposed above can in principle be adopted in the
context of any of the existing calculational techniques, including
those based on the automatic generation of matrix-elements such as
\comphep, \madgraph\ or \grace,
the flexibility to calculate amplitudes relative to fixed colour
states, as well as those relative to dual amplitudes, and the ability
to efficiently calculate matrix elements for processes with large
numbers of final-state partons, single out in our view the \ALPHA\ algorithm as
the best suitable tool.  The ability to tackle computations of this
complexity in the case of hadronic collisions was proved
in~\cite{Caravaglios:1999yr,Draggiotis:1998gr}.

\section{${\mathbf WQ\overline{Q}+n}$ jets production}
\label{sec:notat}                    
The associated production of $W$ bosons and heavy quarks $(Q=b,t)$ is
one of the most important background processes to several searches for
new physics, as well as to the detection of top quarks. The final
state $Wb\bar{b}$ is also the leading irreducible background to the
production of SM Higgs bosons via the process $\ppbar \to
W(H\to b\bar{b})$. The relative matrix elements have been known for
long time~\cite{Kunszt:1984ri}. The NLO corrections to the final
states where both $b$ and $\bar{b}$ are hard and can be treated as
massless have recently become available~\cite{Ellis:1999fv}. The
process $\ppbar \to W b\bar{b} j$ has been studied for massive $b$ at
LO in~\cite{Belyaev:1999dn}, while the processes with up to two extra
jets, but again in the limit of massless $b$, are included in
\vecbos~\cite{Berends:1991ax}.

The extension to larger jet multiplicities, and the inclusion of mass
effects, are necessary, among other things, to allow more accurate
studies of the backgrounds to $t\bar{t}$ production. In this case,
final states have typically 2 jets in addition to the $b\bar{b}$ pair,
but initial and final state radiation can lead to the presence of
extra jets. As we shall show later, the increase in the 
number of light jets in the final state which the background
calculation should be able to cope with is also mandated by the need
to account for background configurations where only one jet can be
reconstructed from the $b\bar{b}$ system.

As an application of our techniques, we then 
carried out the calculation of the process $\ppbar \to
(W\to \ell \nu_{\ell})+Q\overline{Q}+n$-jets, covering final
states with up to $n=4$ jets in addition to the heavy-quark pair.  
The full spin correlations between the leptons from the $W$ decay and
the other particles are included in the matrix element, as well as the
finite width of the $W$, described by a Breit-Wigner. For simplicity,
in the following we shall however always refer to the
$W$, instead of the lepton pair which is used in the actual computations.

The classes of processes necessary to describe these final states, 
and with up to 2 light-quark pairs, are listed here:
\ba
\label{eq:PROC1}
PROC=1: && q \bar{q}' \to W b \bar{b}   \\
PROC=2: && q g \to W b \bar{b} \bar{q}'   \\
PROC=3: && g g \to W b \bar{b} q \bar{q}'   \\
PROC=4: && (q \bar{q}' \to W b \bar{b} q'' \bar{q}'')   + 
           (q \bar{q}'\to W b \bar{b} q \bar{q})   +
           (q \bar{q}'' \to W b \bar{b} q' \bar{q}'')   +
           \\ 
        && (q'' \bar{q}'' \to W b \bar{b} q \bar{q}')   + 
           (q   \bar{q} \to W b \bar{b} q \bar{q}')   + 
           (q   q \to W b \bar{b} q q')   + 
           (q   q' \to W b \bar{b} q q)     \nn \\
PROC=5: && (q g \to W b \bar{b} q' q'' \bar{q}'')   + 
           (q g \to W b \bar{b} q' q \bar{q})   + 
           (q g \to W b \bar{b} q  q \bar{q}')    \\
\label{eq:PROC6}
PROC=6: && (g g \to W b \bar{b} q \bar{q}' q'' \bar{q}'')   + 
           (g g \to W b \bar{b} q \bar{q}' q \bar{q})   \; .
\ea
We did not indicate possible additional final-state gluons, and did
not explicitly list trivial permutations of the initial state partons
and charge-conjugated processes.  This list of processes fully covers
all those possible in the case of up to 3 jets in addition to the $b$
and $\bar{b}$. In the case of 4 extra jets, we did not include
processes with 3 light-quark pairs. Within the uncertainties of the LO
approximation, these can be safely neglected~\cite{Berends:1991ax}. 
The matrix elements are obtained using \ALPHA, which
calculates the Green function generator via an exact
numerical iterative algorithm, as explained in~\cite{Caravaglios:1995cd}.

Note that PROC=6 only contributes to $\ge 4$ light-jet production; 
PROC=5 only contributes to $\ge 3$ light-jet production; 
PROC=3 and 4 only contribute to $\ge 2$ light-jet production; 
PROC=2 only contributes to $\ge 1$ light-jet production;
PROC=1 contributes to all $Wb\bar{b}+X$ final states.
Every time the jet multiplicity is increased, new classes of
processes appear. These new processes cannot be simply obtained by
adding one extra gluon to the lower-order ones. As a result, their
rate cannot be estimated in a shower MC approach, where lower-order
processes are allowed to evolve and emit more jets due to gluon
radiation. This fact stresses once more the importance of a complete
parton-level calculation of all relevant matrix elements.

In the cases where comparisons with existing results were possible, we
verified that our code reproduces previous calculations. We carried
out these tests at the level of total
cross sections for the massive $Wb\bar{b}$ case
(comparing with the results of~\cite{Kunszt:1984ri}), and 
for $Wb\bar{b}$ plus up to 2 extra jets, for massless
$b$ (comparing with the results obtained running the \vecbos\
code\footnote{\tt http://www-theory.fnal.gov/people/giele/vecbos.html}).

All calculations are performed using MC techniques, and the resulting
code can be used as an event generator. For each generated event the
code provides the full kinematics and
flavours of the external partons, selected according
to the relative probabilities, as well as the colour flow. The code
includes an interface to \herwig, allowing the events to be fully
evolved through the coherent parton shower, and to be hadronized. The
full code, as well as more detailed documentation,
are available from the URL:
{\tt http://home.cern.ch/mlm/wbb/wbb.html}.

\section{Study of the partonic results}
\label{sec:parton}
In this section we present some numerical results for parton-level
cross-sections and distributions, using experimental configurations
corresponding to the upgraded Tevatron Collider ($p\bar{p}$ collisions at
$\sqrt{s}=2$~TeV) and to the LHC ($pp$ collisions at
$\sqrt{s}=14$~TeV). 
As default parameters for our calculations we shall use:
$m_W=80.23~\gev$, $\Gamma_W=2.03~\gev$, $m_b=4.75~\gev$, PDF set
CTEQ5M~\cite{Lai:2000wy}, with 2-loop $\as(Q^2)$, and
renormalization/factorization  scales
$\mu_R^2=\mu_F^2=m_W^2+\langle p_T^2 \rangle$, where $\langle p_T^2 \rangle$ is
the average $p_T^2 $ of all outgoing jets. 

We shall define here as light jets those formed by the light quarks
and gluons. They are required to be separated from each other, and
from the $b$ quarks, by $\Delta R_{ij}>0.4$,
with $\Delta R_{ij}= [(\eta_i - \eta_j)^2 + (\phi_i -\phi_j)^2]^{1/2}$
for each pair of  partons $i$ and $j$. The cut at  $\Delta
R_{ij}>0.4$, rather than at the value of  $\Delta R_{ij}>0.7$
used as standard in jet physics at the Tevatron, is motivated by the
choice of jet definition used in most top-quark studies at the
Tevatron~\cite{Abe:1994st}. We shall analyse later the impact of this
choice on the comparison between  jet rates at the parton
level and at the fully-showered level.
Finally, all jets are also required to be
hard and central:
\be \label{eq:ljet}
p_t^i>20~\gev, \quad
\vert \eta_i \vert < 2.5
\ee
We shall not set any cut on the
charged lepton and on the neutrino (giving rise to missing transverse
energy, \met) from the $W$ decay, and we shall present rates including
only one possible leptonic flavour in the $W$ decay. 

In the following, we shall use the symbol $N_J$ to indicate the total
number of jets, including the jets generated by the $b$ or $\bar{b}$
quarks.

\subsection{Total rates}
We start by considering final states where both 
$b$ and $\bar{b}$ are sufficiently hard and
well separated to form independent jets, and apply to them the same
cuts defining light jets, namely $\Delta R_{b\bar{b}}>0.4$ and \eqn{eq:ljet}.
We shall call these ``\twobj\  events''. 
Table~\ref{tab:ratesTEV} gives the rates for  final states with 2 $b$
jets plus extra light jets at
the Tevatron, with the contributions from the different classes of
processes listed separately. The Table shows that, for
multiplicities up to 4 jets, the qualitatively new processes appearing
every time the multiplicity is increased by one are of the same order,
or larger, as the processes obtained from the lower-order channels 
by radiating one extra gluon. 

The results for the LHC are given in Table~\ref{tab:ratesLHC}. Note
here that the effect of the new processes appearing for larger jet
multiplicities is even more important. This is due to the fact that
the lowest-order process involving initial state antiquarks is
strongly suppressed by the sea-quark density relative to higher-order
final states, which are enhanced by the large initial-state gluon
contribution.  This result is consistent with the numbers quoted in
Table~7 of~\cite{Berends:1991ax} in the case of massless $b$.

{\renewcommand{\arraystretch}{1.2}
\begin{table}
\begin{center}
\begin{tabular}{l|lllll} \hline
Process      & $N_J=2$ & $N_J=3$ & $N_J=4$ & $N_J=5$ & $N_J=6$
 \\ \hline \hline
 1      &  360(1)     & 68.6(4)    & 10.4(1)  &  1.46(1)  &  0.20(1)  \\
 2      &  --         & 37.6(2)    & 12.1(1)  &  2.63(3)  &  0.47(1)  \\
 3+4    &  --         & --        &\phantom{1}4.3(1) &   1.66(3)  &  0.41(1) \\
 5      &  --         & --         &  --      &  0.085(2) &  0.036(1)  \\
 6      &  --         &  --        &  --      &  --       &  0.00038(2) \\
\hline
 Total  & 360(1)     & 106.4(4)    & 26.8(2)  &  5.84(4)& 1.11(2) \\
\end{tabular}                                                                 
\ccaption{}{\label{tab:ratesTEV} Contributions from different initial
  states to the $p\bar{p} \to (W\to \ell \nu)
  b\bar{b}j_3\dots j_{N_J} $ rates (fb), for the Tevatron at $\sqrt{S}=2$~TeV,
  with the cuts given
  in eq.~(\ref{eq:ljet}). The different processes 
  are defined in eqs.~(\ref{eq:PROC1})-(\ref{eq:PROC6}). The -- indicates
  that the process is not available for the given jet
  multiplicity. The PDF set is CTEQ5M, and only one lepton flavour is  considered. The uncertainties (quoted in parentheses as errors on
  the last significant figure) reflect the statistical accuracy of our
  MC evaluation. }
\end{center}                                         
\end{table} }

{\renewcommand{\arraystretch}{1.2}
\begin{table}
\begin{center}
\begin{tabular}{l|lllll} \hline
Process      & $N_J=2$ & $N_J=3$ & $N_J=4$ & $N_J=5$ & $N_J=6$
 \\ \hline \hline
 1      & 2.60(1) & 0.63(1) & 0.144(3)   &  0.036(2)  &  0.008(1)\\
 2      &  --     & 2.97(1) & 2.11 (2)   &  1.08(2)   &  0.47(2) \\
 3+4    &  --     & --      & 0.288(1)   &  0.24(1)   &  0.13(2) \\
 5      &  --     & --      &  --        &  0.030(1)  &  0.031(4) \\
 6      &  --     &  --     &  --        &  --        &  0.0010(3) \\
\hline
 Total  & 2.60(1) & 3.60(1) & 2.54(2)    &  1.38(2)   & 0.64(3) \\
\end{tabular}                                                                 
\ccaption{}{\label{tab:ratesLHC} Same as Table~\ref{tab:ratesTEV},
  for the LHC. Rates in pb. }
\end{center}                                         
\end{table} }

Since experimentally one does not always require the identification of
both $b$ and $\bar{b}$, it is important to consider, in addition to the
case of \twobj\  events, cases where the event has only one
taggable \bjet.
These events receive contributions from final states where 
either one between the $b$ and $\bar{b}$ is too
soft or outside the rapidity range for the jet definition, or where 
the $b$ and $\bar{b}$ are close enough as to merge into a single jet.
If we treated the $b$ as a massless particle, 
both these limiting cases would lead to infinite rates at LO,
due to soft or collinear divergences. 
Only the inclusion of NLO virtual corrections, with an infrared and
collinear safe jet definition, would restore a finite and physical
answer. Since we treat the $b$ quark as massive, the LO calculation is however
meaningful and finite 
(expressing the fact that the ``$b$-ness'' of a jet is in
principle measurable via its decays regardless of how small its
energy is, and regardless of how collinear the $b$-$\bar{b}$ pair is). 

We can therefore define and 
predict the rates for final states where only one jet is taggable. 
We call these ``\onebj\  events''.
More specifically, 
\onebj\  events are those which fail the \twobj\  definition,
but fall in one of these classes:
\begin{enumerate}
\item One between the $b$ and the $\bar{b}$ satisfies \eqn{eq:ljet}.
\item Both $b$ and $\bar{b}$ fail \eqn{eq:ljet}, but
 \be
 \Delta R_{b\bar{b}}<0.4, \;
 p_t(\vec{b}+\vec{\bar{b}})>20~\gev, \; \vert
  \eta(\vec{b}+\vec{\bar{b}}) \vert < 2.5 \; .
  \ee
\end{enumerate}
For these events, production of a $W+N_J$-jet final state requires an
${\cal{O}}(\alpha_s^{N_J+1})$ process, as one extra light-parton jet
is needed to give the required jet multiplicity. In spite of the presence
of the extra power of $\alpha_s$, the  contribution of these events to
the 1-tag final states can be very large, because of the presence of
potentially large soft and
collinear logarithms.  This is confirmed by Table~\ref{tab:bjetsTEV}, which
shows that the contribution to single-tag events coming from
the higher-order processes is as large as that of the leading-order 
ones\footnote{ The actual relative
  contribution of \twobj\  events and \onebj\  events to the
  single-tagged rate depends on the experimental value of the tagging
  efficiency, which will depend on the structure of the $b$ jet. We
  expect, for example, that tagging efficiencies for semileptonic tags
  will be approximately a factor of two larger for jets containing two
  $b$ than for single-$b$ jets.}.
In the case of the LHC, Table~\ref{tab:bjetsLHC}, the higher-order
contributions are significantly larger than the leading-order ones,
consistently with the results shown for the multiplicity-dependence of
the total jet rates in Table~\ref{tab:ratesLHC}. This is particularly
true for the process most relevant to the study of associated $WH$
production, where the ${\cal{O}}(\alpha_s^{3})$ process is
almost 5 times larger than the
${\cal{O}}(\alpha_s^{2})$ one. The requirement of double-$b$ tagging
is very important in this case to efficiently reduce this extra background.

Before concluding this section we stress that all rates shown in the
Tables are affected by a large overall uncertainty due to the choice of
renormalization and factorization scales. Since the matrix-element
calculations only include the LO contributions, and given the large
powers of $\as$ which multiply the rates (the processes with $N_J$
final state partons scale like $\as^{N_J}$), one should assume overall
systematic uncertainties of the order of a factor of 2-4, depending on
the jet multiplicity. In principle, this uncertainty can be reduced by
using the information coming from data on $W+N_J$ jets (without
final-state \bjet s) or $Z++b\bar{b}+n$ jets, which are not
affected by the contamination of the $t\bar{t}$ signal. The
calculation of the $Z+b\bar{b}+n$~jet processes will be done in the future.

{\renewcommand{\arraystretch}{1.3}
\begin{table}
\begin{center}
\begin{tabular}{l|lllll} \hline
Perturbative Order           & $N_J=1$ & $N_J=2$ & $N_J=3$ & $N_J=4$& $N_J=5$
 \\ \hline \hline
 ${\cal{O}}(\alpha_s^{N_J})$   & --      & 360(1) & 106.4(4)  &
 26.8(2)& 5.84(4) \\ \hline
 ${\cal{O}}(\alpha_s^{N_J+1})$ & 1316(3)& 371(2) & \phantom{1}94.0(5) &
 20.5(4)&  3.8(1)
\end{tabular}                                                                 
\ccaption{}{\label{tab:bjetsTEV} Contributions to $W+N_J$-jet rates (fb)
 at the Tevatron, from
 final states with 2 (upper row) or 1 (lower row) $b$-jets. }
\end{center}
\end{table}}

{\renewcommand{\arraystretch}{1.3}
\begin{table}
\begin{center}
\begin{tabular}{l|lllll} \hline
Perturbative Order           & $N_J=1$ & $N_J=2$ & $N_J=3$ & $N_J=4$& $N_J=5$
 \\ \hline \hline
 ${\cal{O}}(\alpha_s^{N_J})$   & --     & \phantom{1}2.60(1)    & 3.60(1) &
 2.54(2) & 1.38(2) \\
\hline
 ${\cal{O}}(\alpha_s^{N_J+1})$ & 9.38(5) & 12.3(1) & 7.4(1) & 3.71(5)
 & 1.7(1)
\end{tabular}                                                                 
\ccaption{}{\label{tab:bjetsLHC} Contributions to $W+N_J$-jet rates (fb)
 at the LHC, from
 final states with 2 (upper row) or 1 (lower row) $b$-jets. }
\end{center}
\end{table}}

\subsection{Jet distributions}
\begin{figure}
\begin{center}
\includegraphics[width=0.49\textwidth,clip]{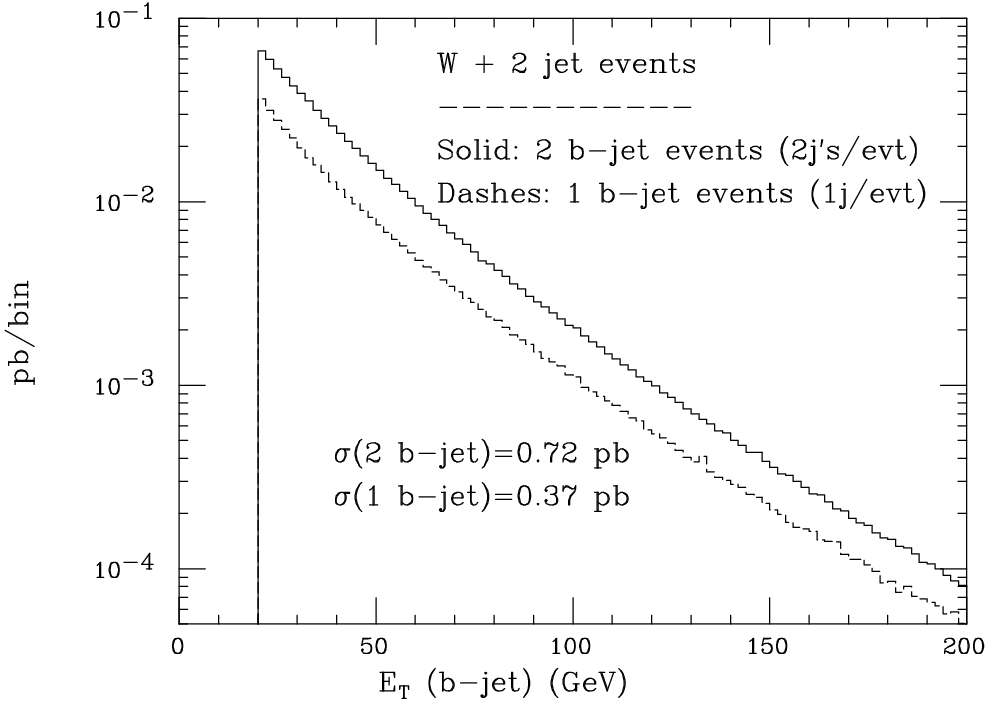} \hfill
\includegraphics[width=0.49\textwidth,clip]{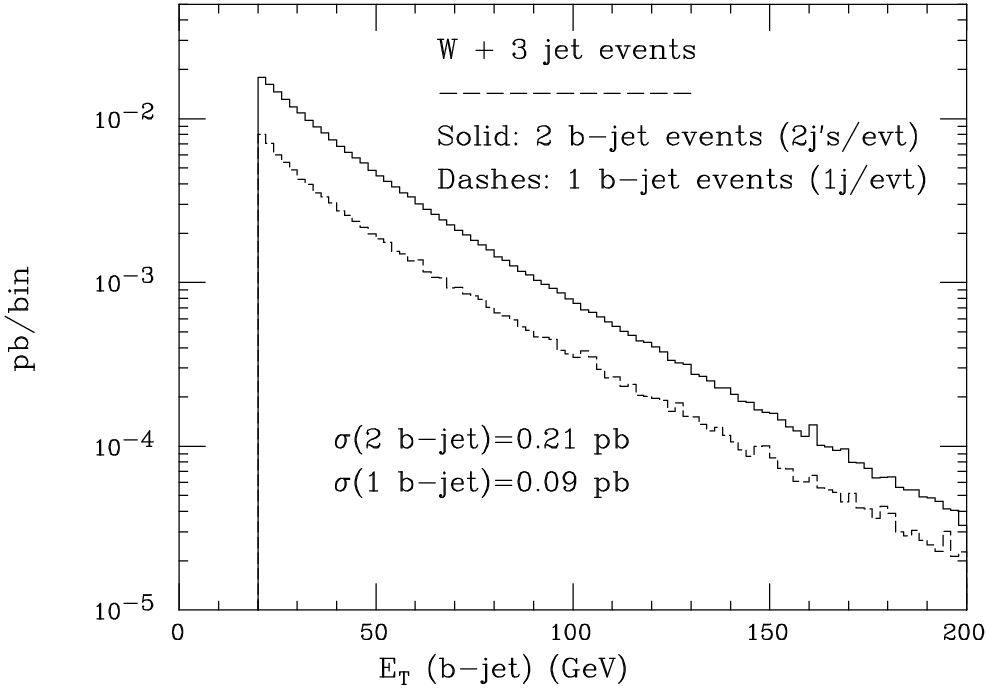} \\
\includegraphics[width=0.49\textwidth,clip]{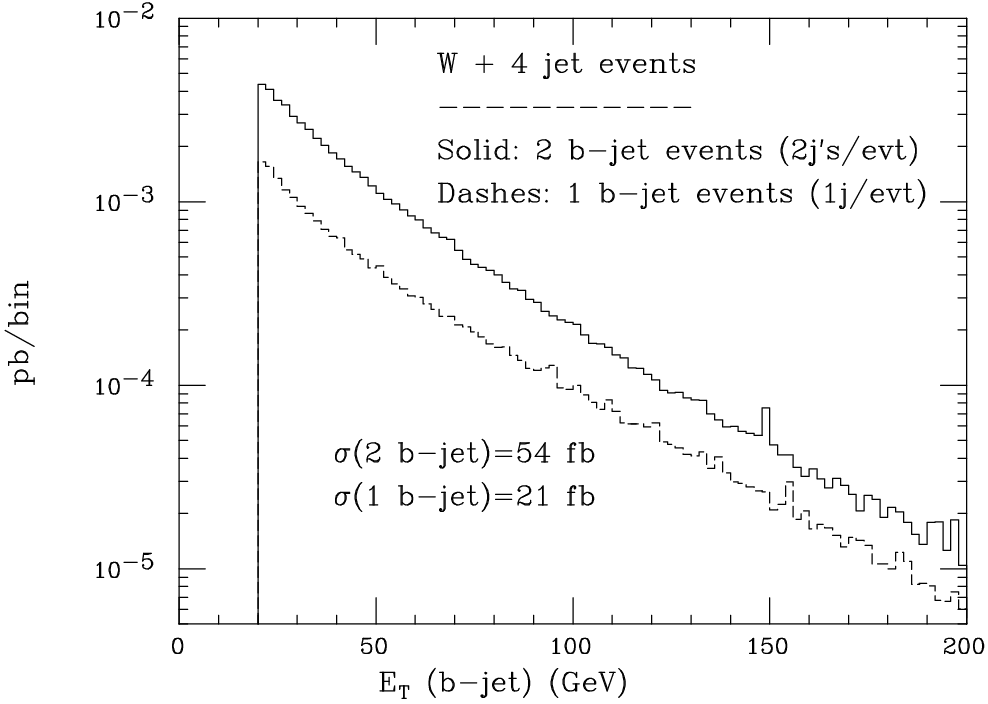} \hfill
\includegraphics[width=0.49\textwidth,clip]{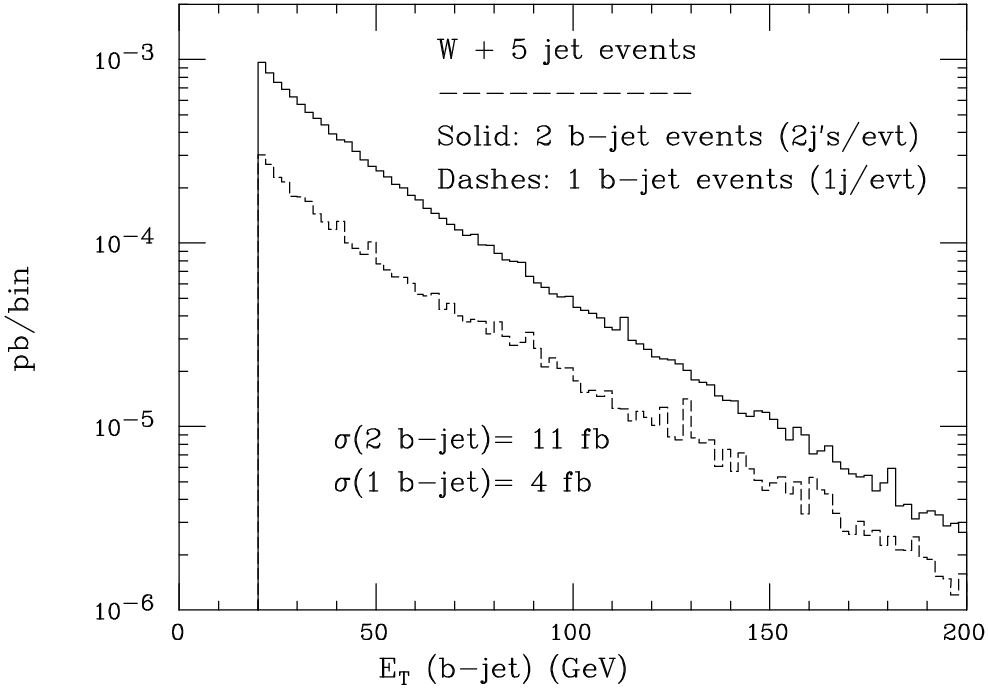}
\ccaption{}{\label{fig:2jbpt} Inclusive $\pt$ distributions of $b$
  jets. The solid lines refer to \twobj\  events (in which case both
  jets are entered in the histograms); the dashed lines refer to 
   \onebj\  events.}
\end{center}
\end{figure}

We present in this subsection some interesting distributions for parton-level
multijet events. Figure~\ref{fig:2jbpt} shows the inclusive $\pt$
distribution of $b$ jets, for final states with  jet
multiplicities $N_J=2$-5. We compare the distributions of 
$b$ jets in \twobj\ and \onebj\ events, where in the first case each
event will contribute two entries to the histograms. The shapes of the
two distributions are rather similar through most of the energy
range. The curve relative to \twobj\ events has a larger normalization
because of the double probability of finding a \bjet\ in the event.

\begin{figure}
\begin{center}
\includegraphics[width=0.49\textwidth,clip]{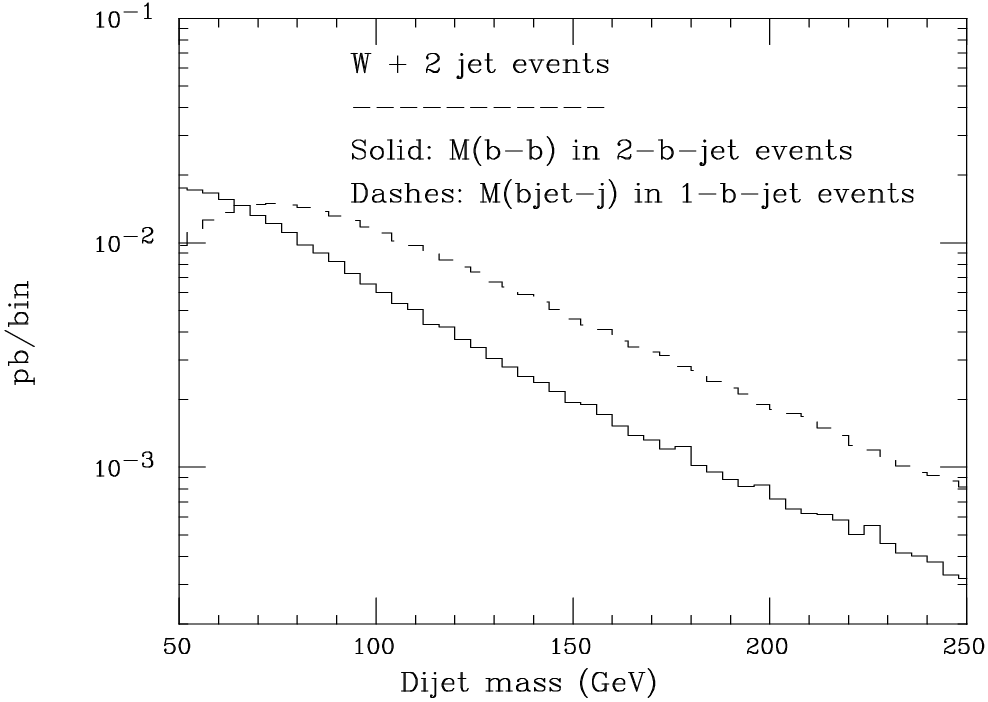}
\ccaption{}{\label{fig:bbmass} Dijet invariant mass distribution, in
  $W$ plus 2 jet events, where at least one of the two jets has a $b$
  in it. The solid line refers to \twobj\  events; the dashed line refers to 
   \onebj\  events.}
\end{center}
\end{figure}

The large contribution coming from events where only one $b$ jet is
reconstructed is clearly visible. To show the impact that such events
have on the background to the associated production of a $W$ boson and
a resonance decaying to a $b$ pair (e.g. a Higgs boson), we present in
fig.~\ref{fig:bbmass} the invariant mass spectrum of the dijet pair in
dijet events. The solid line corresponds to the \oatwo\ process, where
both $b$ quarks give rise to independent jets. The dashed line
corresponds to the \onebj\ events obtained from the \oacube\
process. This figure shows that the requirement of having two
independently $b$-tagged jets is crucial for the background
suppression. A detailed study of the tagging efficiencies for 
\onebj\ events is necessary to ensure that the residual contribution
from fake tags on the second, non-$b$, jet is small.

\begin{figure}
\begin{center}
\includegraphics[width=0.49\textwidth,clip]{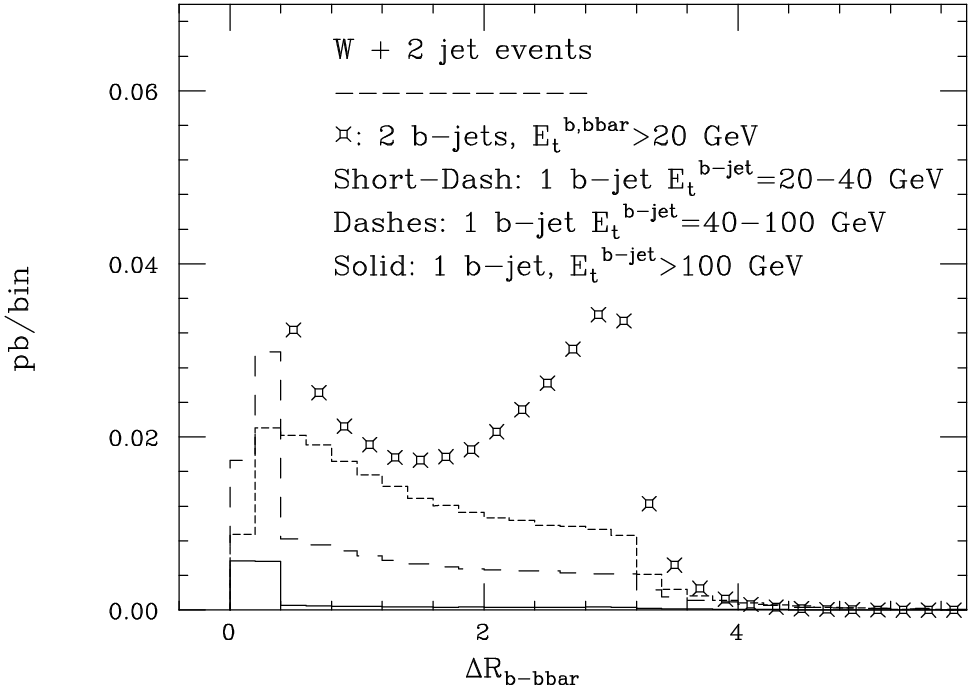} \hfil
\includegraphics[width=0.49\textwidth,clip]{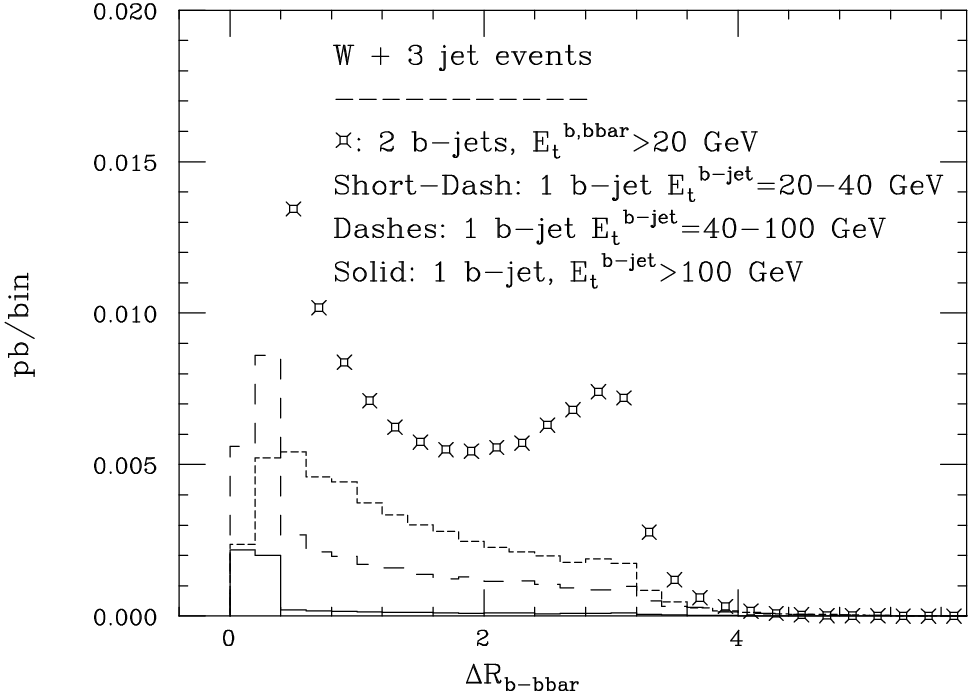} \\
\includegraphics[width=0.49\textwidth,clip]{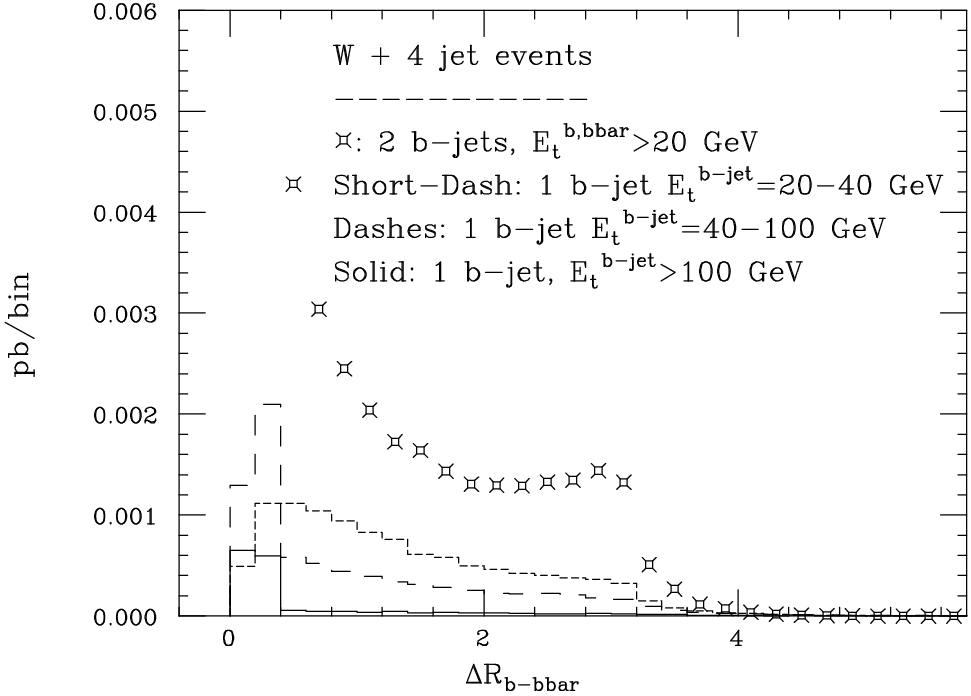} \hfil
\includegraphics[width=0.49\textwidth,clip]{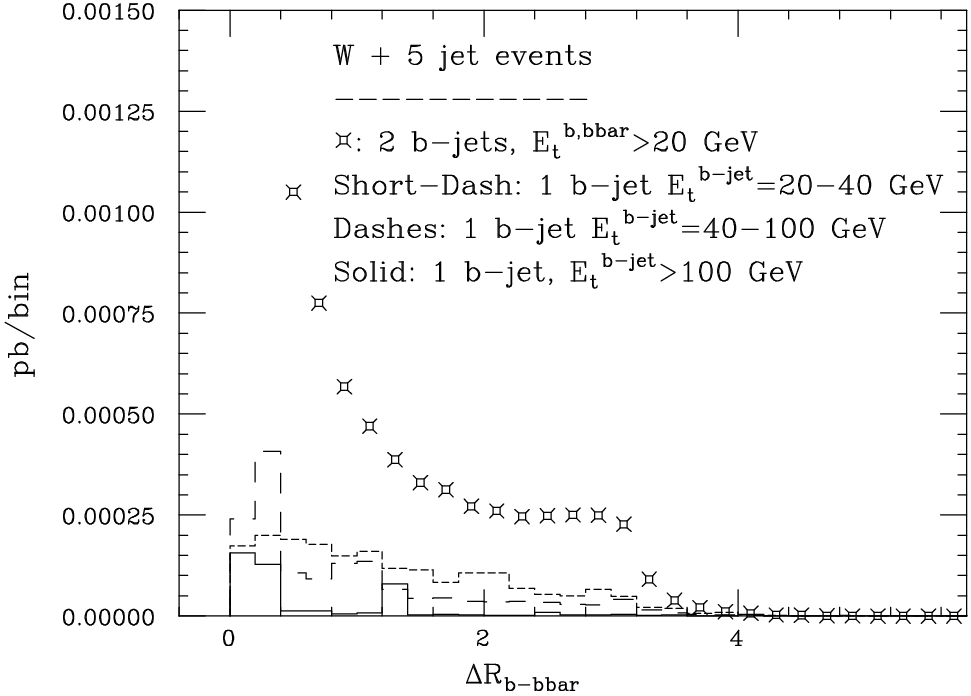}
\ccaption{}{\label{fig:2jrbb} $b\bar{b}$ angular correlations in
  events with 2 $b$ jets (plots) and with 1 $b$ jet (histograms). The case of 1
  $b$ jet is divided into subsamples defined by the $\pt$ of the $b$
  jet in the ranges 20-40, 40-100 and $>100~\gev$.}
\end{center}
\end{figure}

To explore more
in detail the structure of the \bjet\ in \onebj\ events, we plot in
fig.~\ref{fig:2jrbb} the
$\Delta R$ separation between the $b$ and the $\bar{b}$, showing the
curves for different ranges of $\pt$ of the \bjet. Events where
$\Delta R>0.4$ correspond to cases where one of the two $b$s is either
too soft or is outside the rapidity range. The figures show that at
large \pt\ we are dominated by \onebj s with
the   $b$ and $\bar{b}$ merged inside the jet ($\Delta R<0.4$). 
The tagging efficiency for these jets is
 presumably larger than that for jets made of a single 
 $b$, in particular if semileptonic tags are considered. It is an
 important experimental issue to evaluate what the actual tagging
 efficiency is for these jets, as a function of $\pt$.

\section{Study of the fully showered final states}
\label{sec:shower}
In this Section we describe the results obtained after evolution
through the parton shower. 
The goal of our calculations is to be able to generate as accurately
as possible the full jet structure of the partons generated at the
matrix-element level. We first generate a sample of parton-level
events with a fixed multiplicity ($N_J$), and then process these events
through \herwig\ for the shower evolution.
When applied to an event with $N_J$ hard
final-state partons, we expect that the parton-shower evolution will
generate an event with an $N_J$-jet inclusive final state. Initial or
final state radiation may change the overall jet multiplicity (e.g. by
splitting a jet in two, or radiating a new one from the initial
state), but these effects are of order $\as$ relative to the inclusive
$N_J$-jet properties of the event. In principle one could use the shower
evolution to predict the rates for configurations with jet
multiplicities larger than $N_J$~\cite{Giele:1990vh,Benlloch:1992fk}. 
We shall give examples later on of how well the MC succeeds in this
goal. As is well known, however, the shower MCs tend to underestimate
the fraction of events with extra jets, unless explicit matrix-element
corrections for the emission of hard and non-collinear radiation are
included. Algorithms exist for implementing these matrix-element
corrections in the case of low jet multiplicity processes (corrections
to DY production~\cite{Seymour:1995df}, top
decays~\cite{Corcella:1998rs}, $WZ$-pair
production~\cite{Dobbs:2001gb}), but their extension to the case of
high-jet multiplicities we are interested in is currently severely
limited by the complexity of these processes.

In the following two subsections we present some tests of our scheme.
In the first one we discuss some sanity checks of our approach,
showing that the inclusive properties of $N_J$-jet final states are
preserved by the shower evolution. In the second subsection we discuss
the ability of the shower MC evolution to predict the emission rates
for extra jets.

\begin{figure}
\begin{center}
\includegraphics[width=0.49\textwidth,clip]{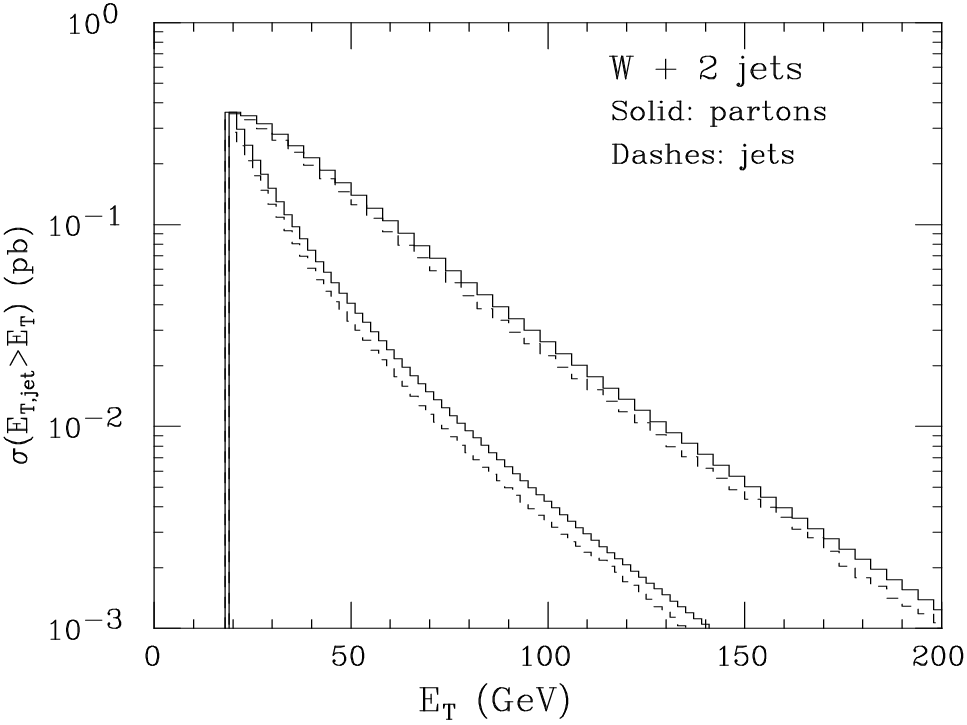}
\hfill
\includegraphics[width=0.49\textwidth,clip]{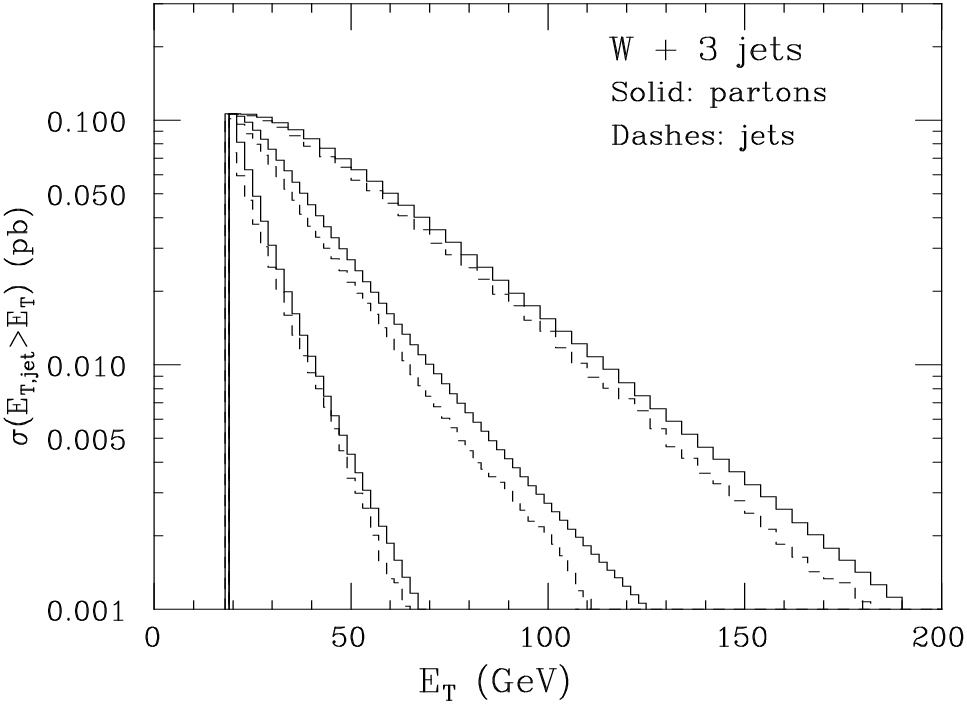}
\\
\includegraphics[width=0.49\textwidth,clip]{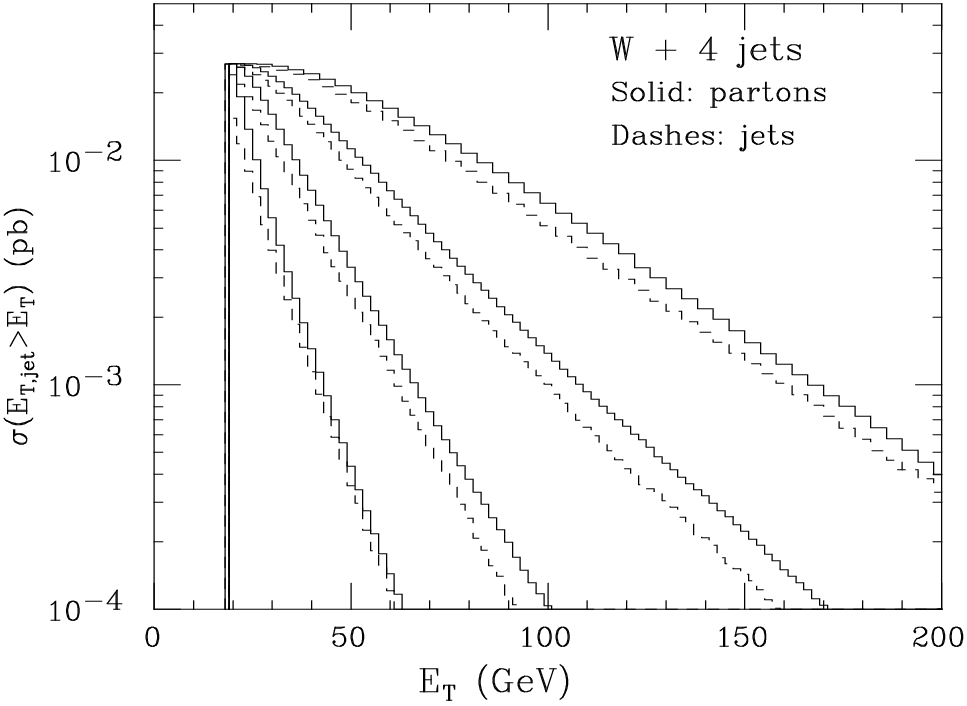}
\hfill
\includegraphics[width=0.49\textwidth,clip]{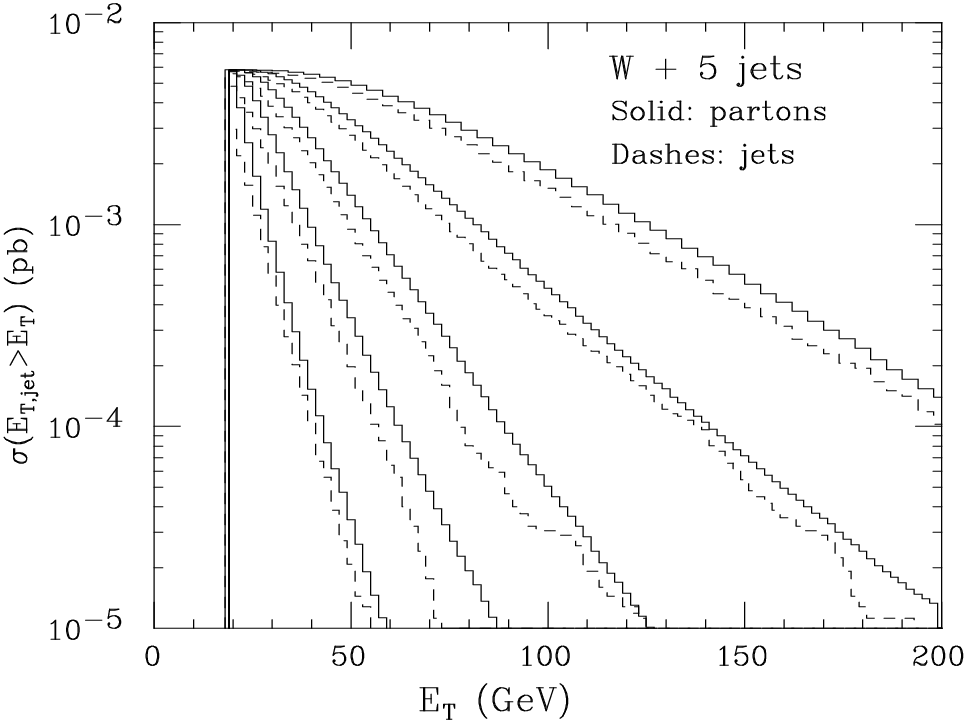}
\ccaption{}{\label{fig:2jhwpt} Inclusive $\pt$ distributions of jets
  at the parton level (solid curves) and after shower evolution
  (dashed curves). The curves are ordered going from the hardest to
  the softest jet in the event. Parton level jets are defined by an
  isolation cut $\Delta R>0.4$, and showered jets by a cone $\rjet=0.4$.}
\end{center}
\end{figure}

\begin{figure}
\begin{center}
\includegraphics[width=0.45\textwidth,clip]{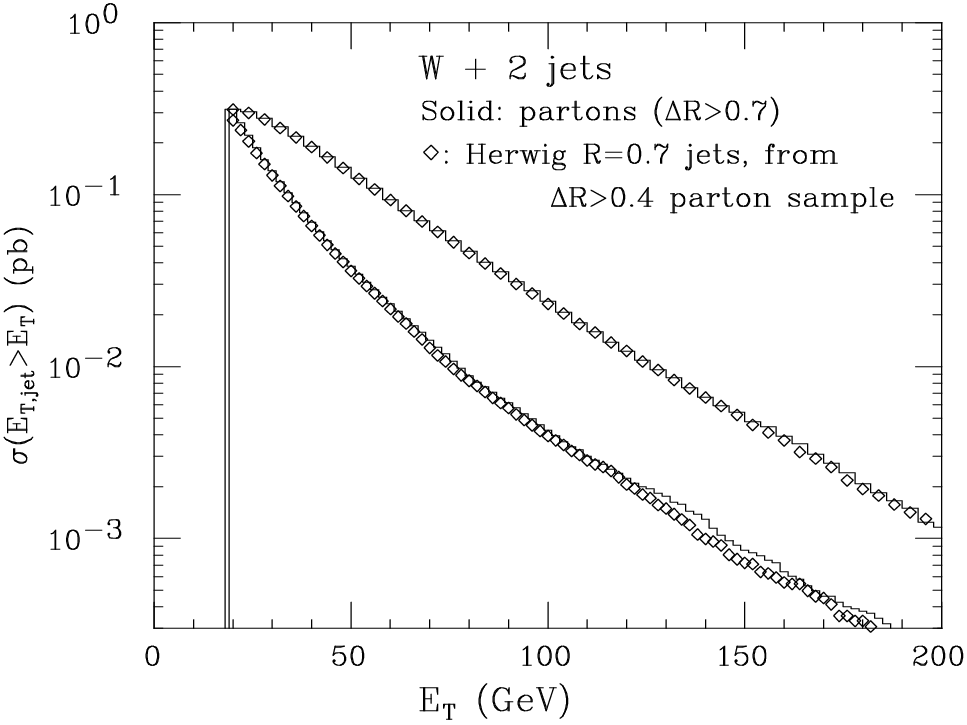}
\hfill
\includegraphics[width=0.45\textwidth,clip]{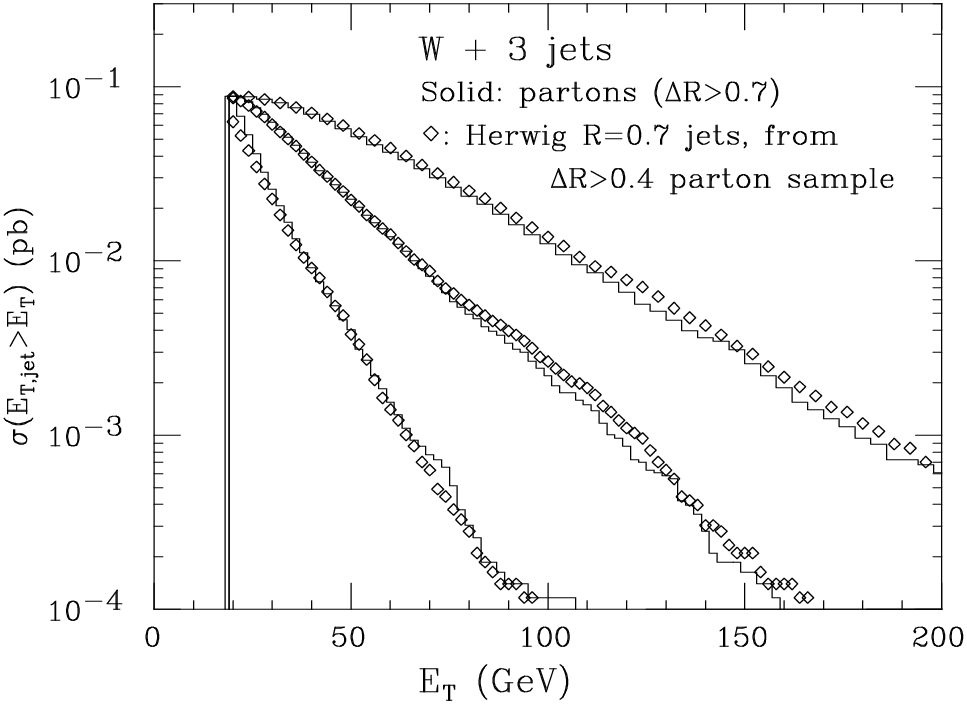}
\ccaption{}{\label{fig:3jpt07} Inclusive $\pt$ distributions of jets
  at the parton level, with separation cut $\Delta R>0.7$ (solid
  curves), and of fully-showered $\rjet=0.7$ jets   (dashed curves). 
  These last were
  obtained starting from the full sample of $\Delta R>0.4$ partonic
  events.}
\end{center}
\end{figure}

\subsection{Sanity checks}
To start with, we compare the inclusive jet rates before and after
evolution.
While the evolution of the partonic events
through \herwig\ can generate fully hadronized final states, we chose
here to stop the evolution after its perturbative part, in order to be
able to compare as closely as possible the  effects of higher-order
perturbative corrections to the evolution of the final state.
For the jet clustering after the
shower evolution we use the
standard cone algorithm, implemented in the routine {\small
  GETJET}~\cite{getjet} provided with the \herwig\ code. 
To match the cuts used at the parton level,
we use
$\rjet=\Delta R=0.4$,
and  a jet $\pt$ threshold of 20~GeV. 

Figure~\ref{fig:2jhwpt} compares the 
$p_T$ spectra at parton level with those after the shower, for events
generated at the parton level with 2, 3, 4 and 5 jets.
The pair of curves
correspond to the series of jets, from the hardest to the softest
one. The curves show  that the spectra
after radiation do not exactly coincide with those at the parton
level. The difference is consistent with an average energy-loss of 3-4~\gev\
 outside the jet cone. One should therefore expect that a better
 matching of the partonic and fully showered jet spectra  will be
 achieved by using wider jet cones, since these will more efficiently
 contain the energy radiated by the partons during their evolution. 
 That this is the case, is shown in
 fig.~\ref{fig:3jpt07}. Here we compare the spectra obtained using a
 cut of $\Delta R>0.7$ on the partons with those relative to fully
 showered jets defined by a cone of $\rjet=0.7$. 
  To generate this sample of $\rjet=0.7$ jets we used  the sample of
 parton-level events defined by the $\Delta R>0.4$ cut, in order to
 cover the cases where two partons merge into a single jet, and extra
 jets are produced by the radiative processes. The agreement between
 the spectra is now perfect, consistently with the results obtained in
 the 2-jet case in~\cite{Giele:1990vh}.
 The message here is that when preparing parton-level samples to be
 used for the QCD evolution by a shower MC algorithm, it is important
 to set generation cuts looser than those used in the definition of
 the full jets, in order to account for downward fluctuations in the
 jet energy and for jet merging induced by the jet clustering
 algorithms. In principle these problems should be avoided by using
 jet algorithms based on $k_T$ clustering~\cite{Catani:1992zp}.


\begin{figure}
\begin{center}
\includegraphics[width=0.45\textwidth,clip]{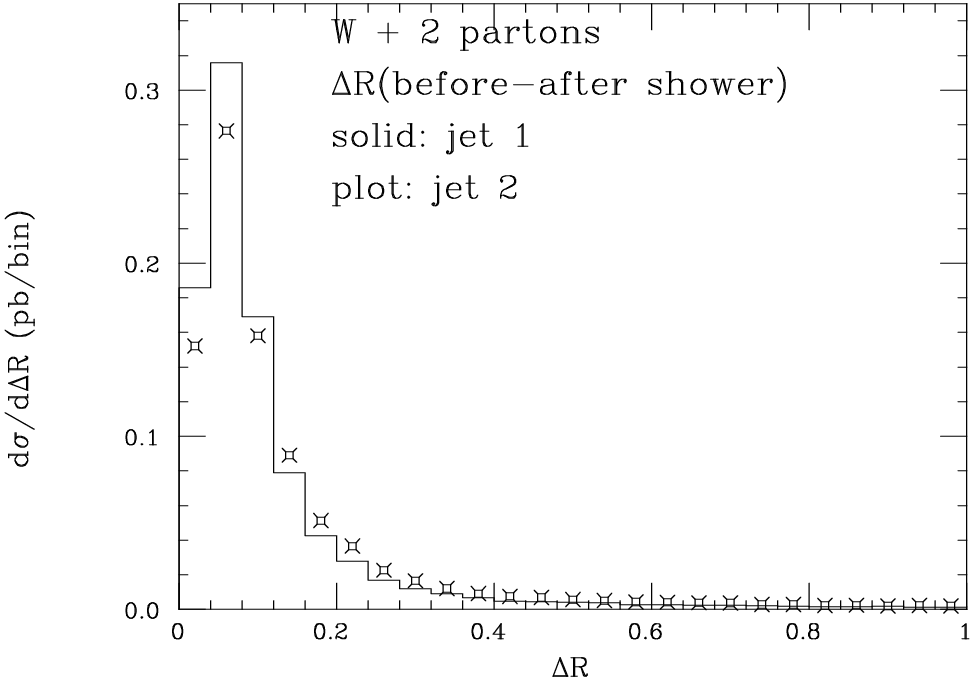}
\hfill
\includegraphics[width=0.45\textwidth,clip]{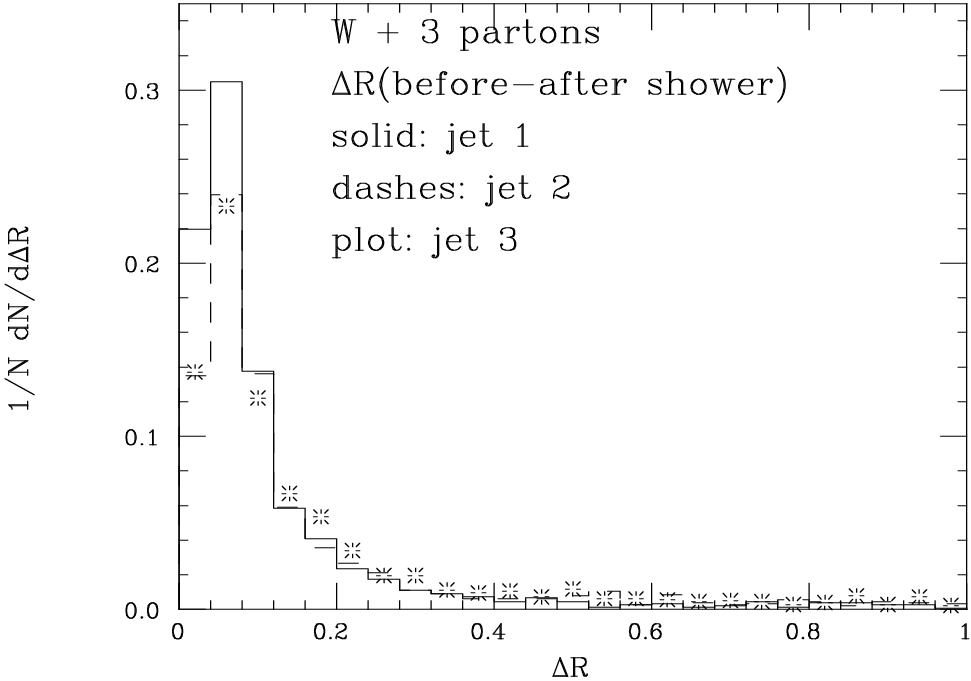}
\\
\includegraphics[width=0.45\textwidth,clip]{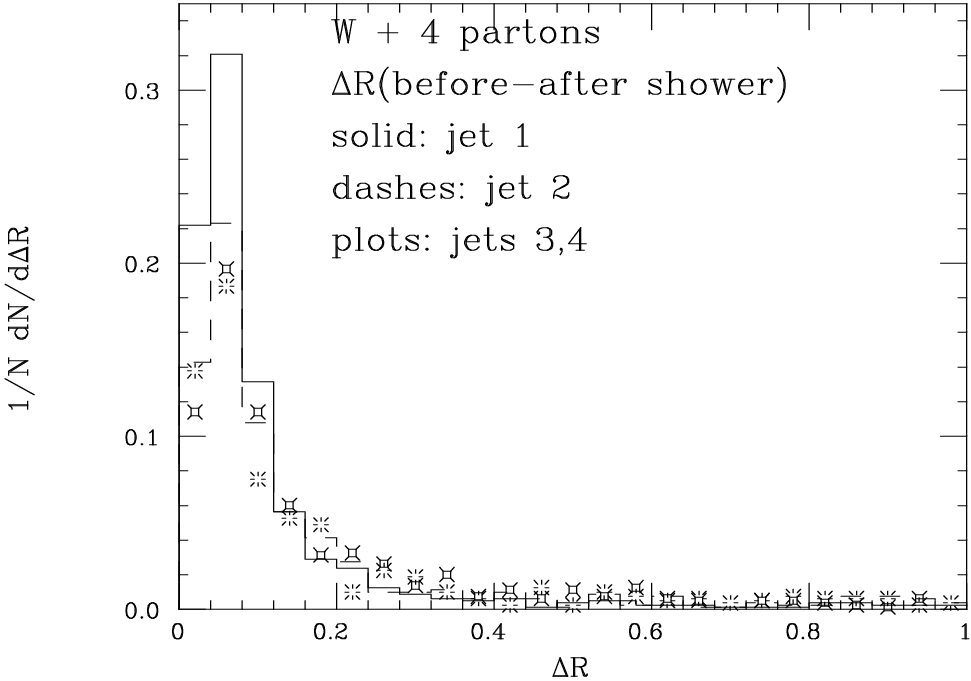}
\hfill
\includegraphics[width=0.45\textwidth,clip]{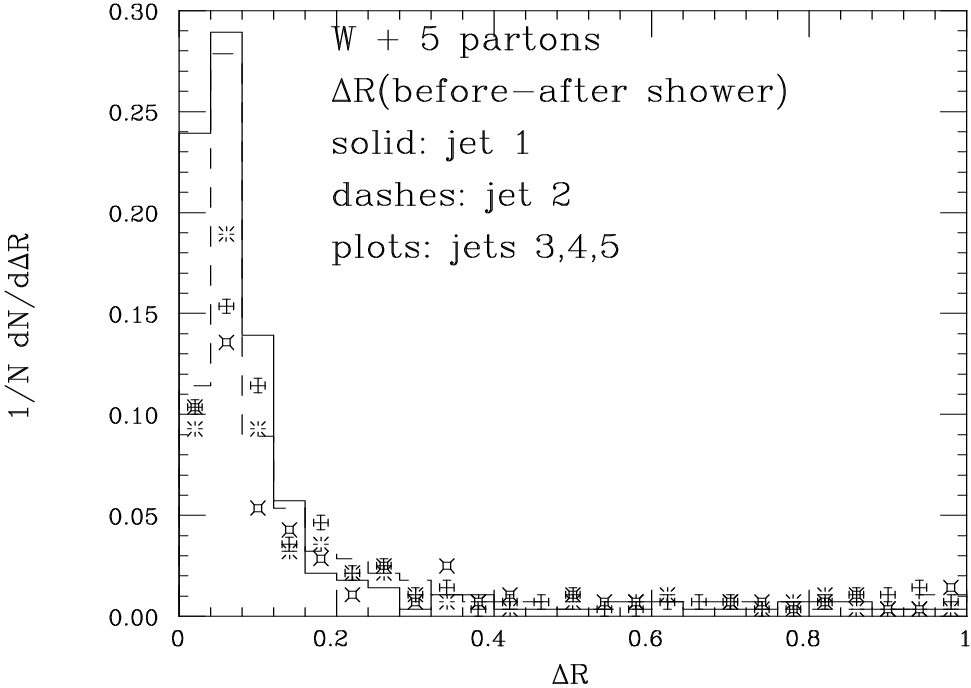}
\ccaption{}{\label{fig:2jhwdRj} Parton-jet alignment.
  We plot the distance $\Delta R$ between the direction of the parton
  before radiation, and that of its daughter jet. Jets are labeled in
  decreasing order of \pt.}
\end{center}
\end{figure}

\begin{figure}
\begin{center}
\includegraphics[width=0.45\textwidth,clip]{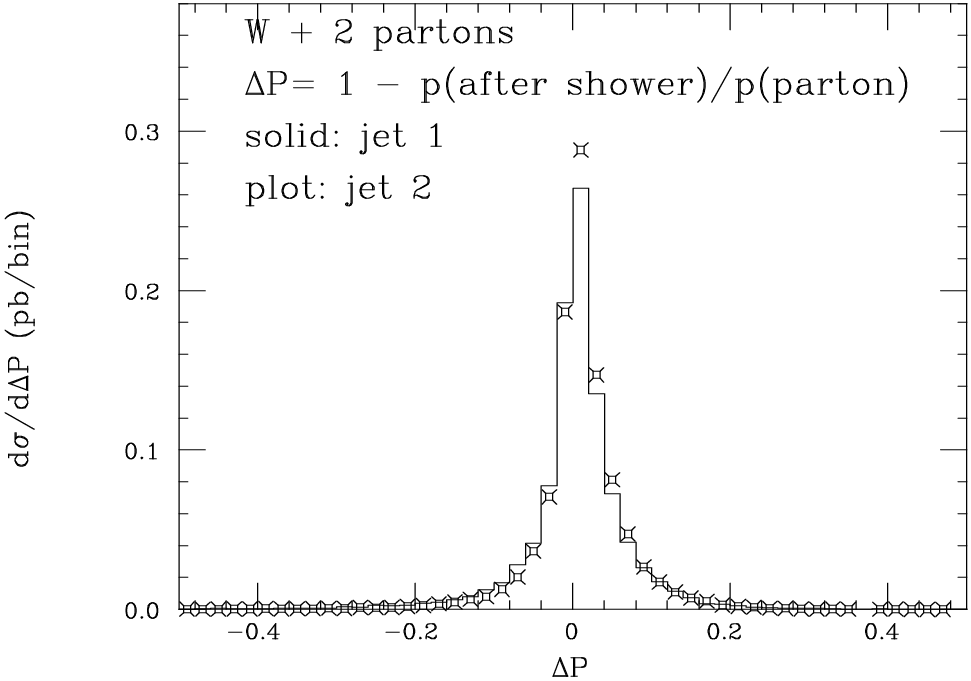}
\hfill
\includegraphics[width=0.45\textwidth,clip]{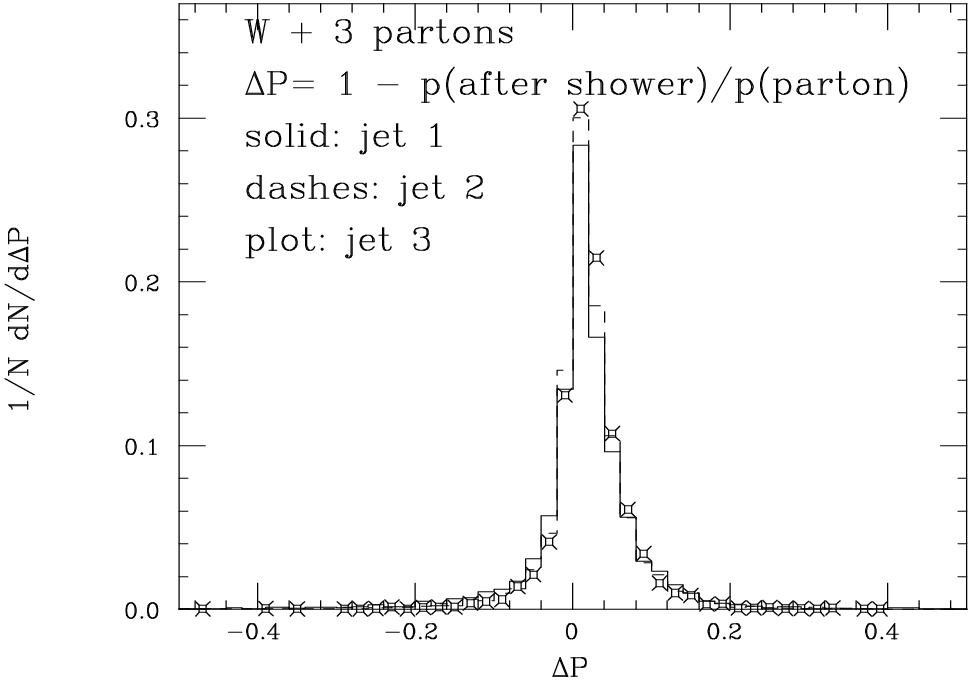}
\\
\includegraphics[width=0.45\textwidth,clip]{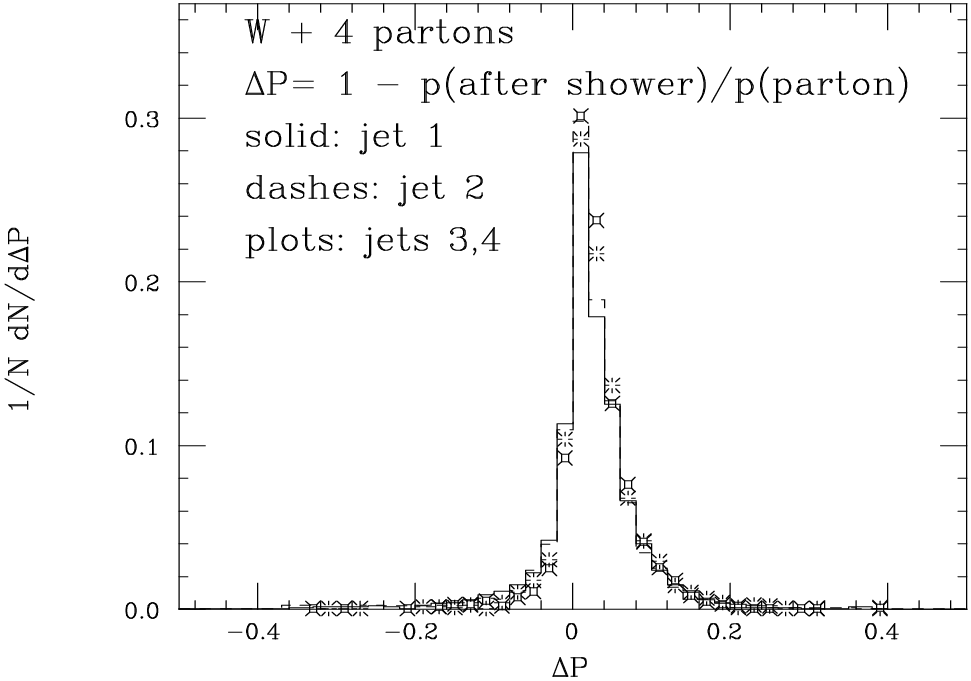}
\hfill
\includegraphics[width=0.45\textwidth,clip]{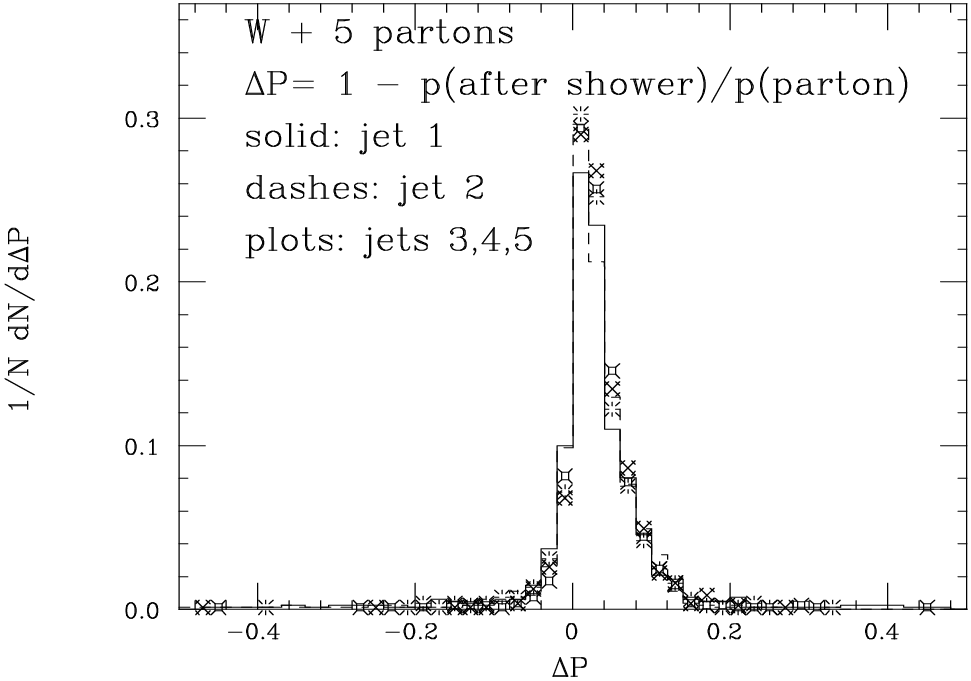}
\ccaption{}{\label{fig:2jhwdpt} Jets' momentum matching. 
  We plot the difference between the momentum of the parton before radiation
  and that of the jet, normalized to the parton momentum.
  Jets are labeled in
  decreasing order of \pt.}
\end{center}
\end{figure}

Figure~\ref{fig:2jhwdRj} shows the matching between the directions of
the partons before the shower, and the direction of the fully evolved
jets. We note a smearing of the direction, which is more enhanced in
the case of the soft jets, as should be expected. The matching
of the jet momenta before and after the shower is shown in
fig.~\ref{fig:2jhwdpt}. The distributions show on average a momentum
loss, mostly due to radiation outside the jet cone, but have a width
which is only of the order of 5-10\% of the parton momentum.

\begin{figure}
\begin{center}
\includegraphics[width=0.45\textwidth,clip]{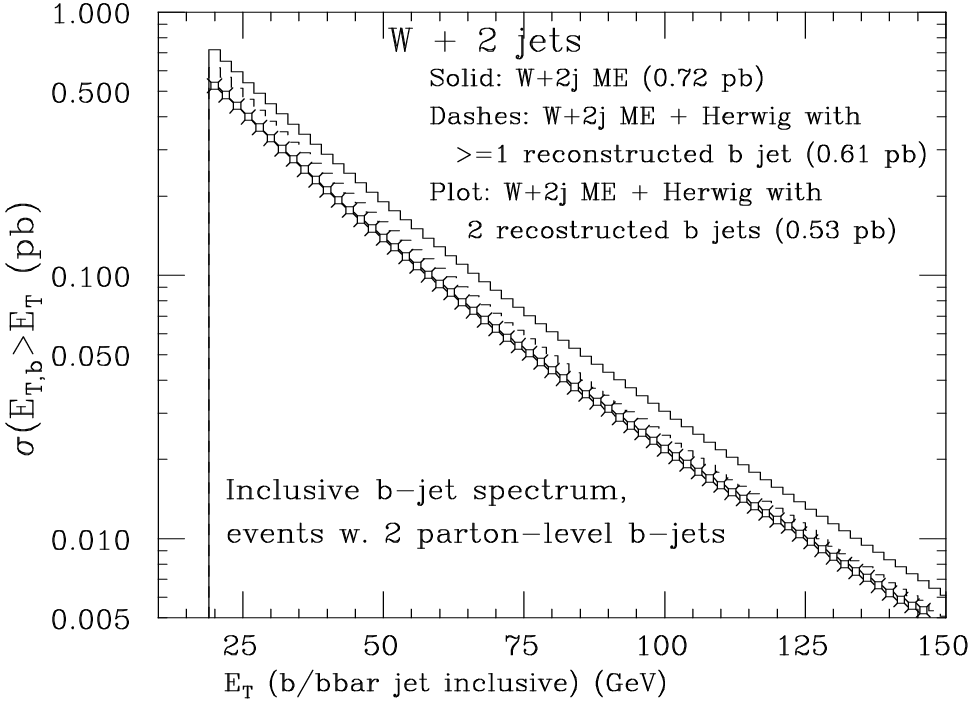}
\hfill
\includegraphics[width=0.45\textwidth,clip]{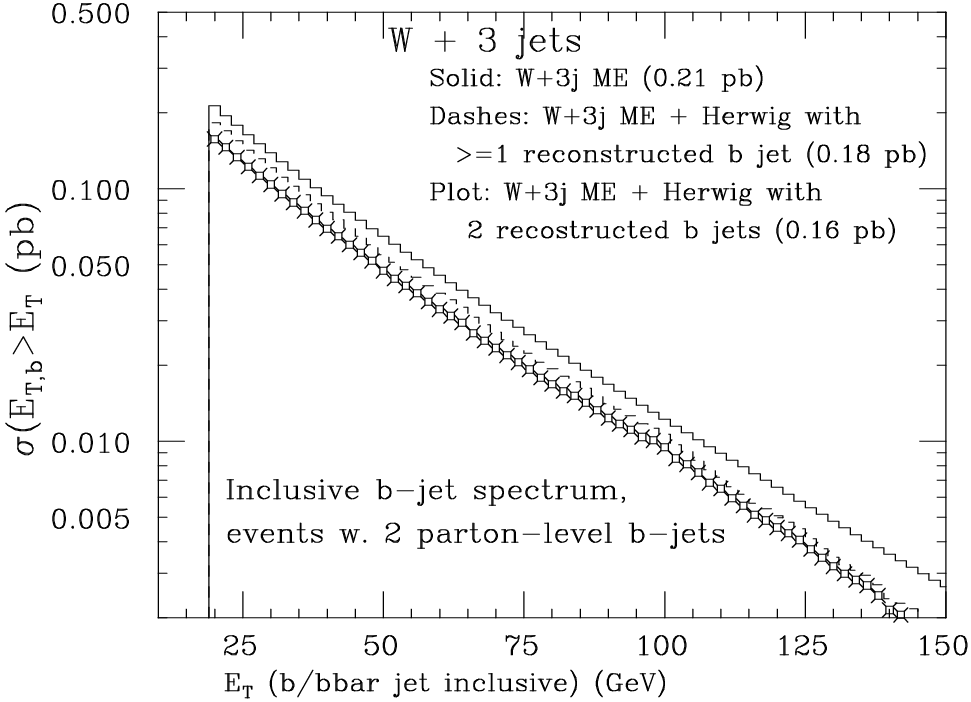}
\\
\includegraphics[width=0.45\textwidth,clip]{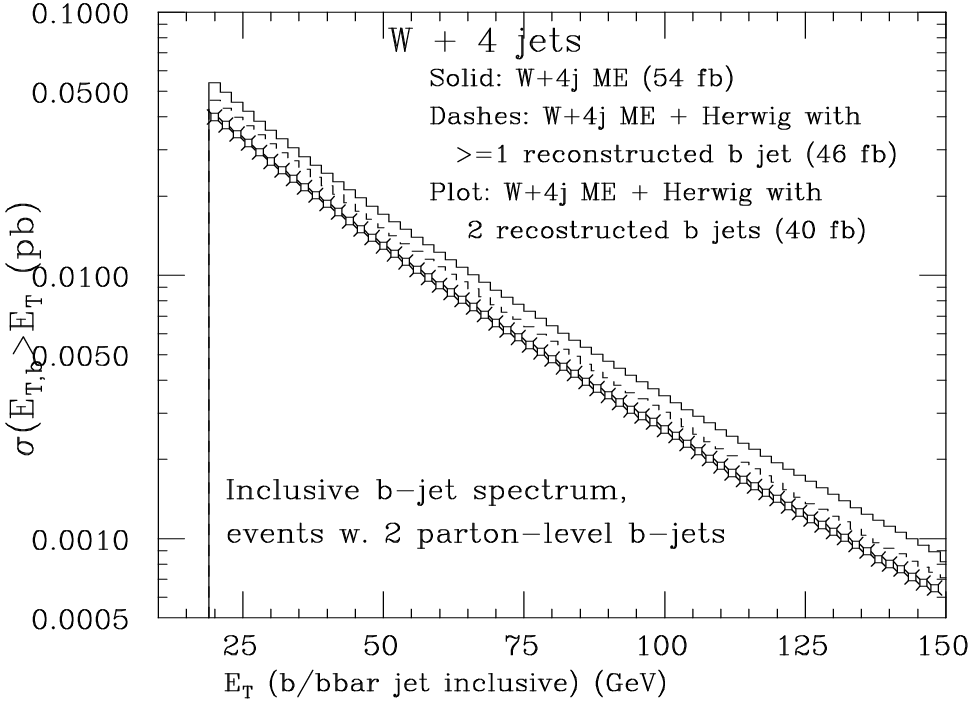}
\hfill
\includegraphics[width=0.45\textwidth,clip]{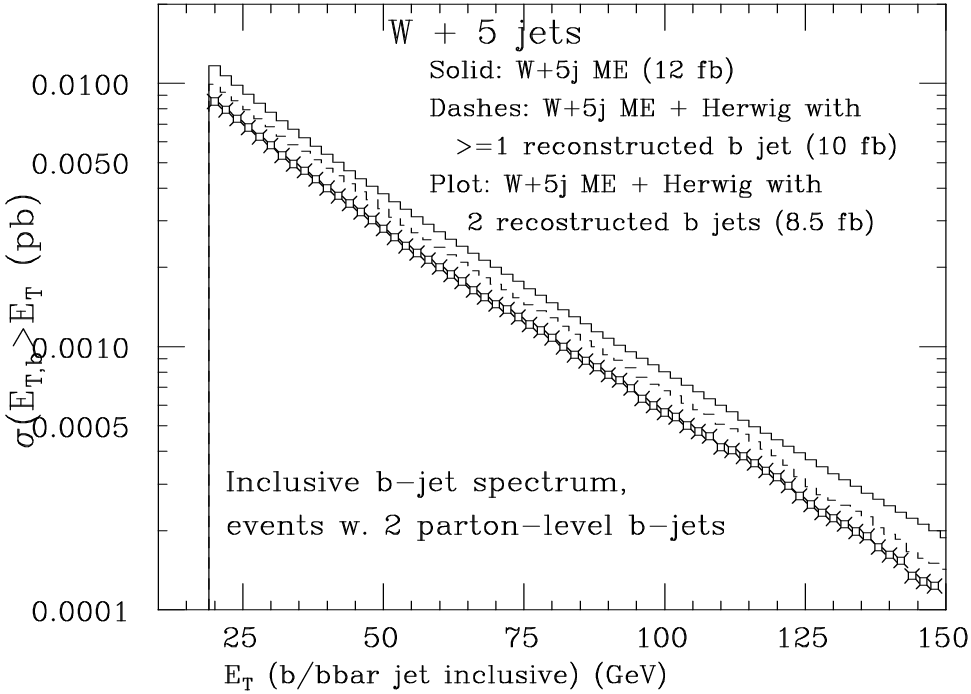}
\ccaption{}{\label{fig:2jhwptb} 
  Inclusive spectrum of $b$ jets: the result for \twobj\ parton-level
  events (solid) is compared to the spectrum obtained for the same
  events after shower evolution (dashed line); 
  the plotted points correspond to the result for 
  events which have still 2 $b$ jets after the shower evolution).}
\end{center}
\end{figure}

In fig.~\ref{fig:2jhwptb} we display the comparison between the inclusive
spectra of \bjet s, before and after shower evolution. At the parton
level we use events with \twobj s. Only a fraction of them will
survive as \twobj\ events after the shower, once again due to some
energy loss outside the cone, and to artifacts of the jet merging
during the clustering. 

The outcome of all these studies is that the shower evolution
preserves the inclusive properties of parton-level events, up to
corrections induced by energy losses outside the small, $\rjet=0.4$,
jet cones. 


\subsection{Jet radiation in \herwig}
\begin{figure}
\begin{center}
\includegraphics[width=0.45\textwidth,clip]{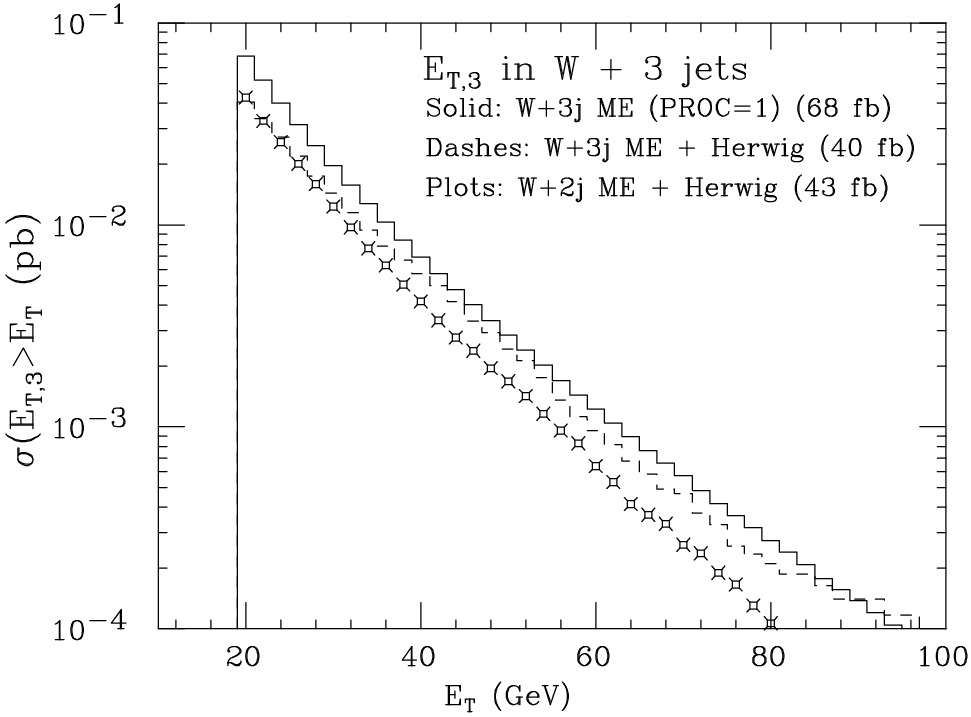}
\hfill
\includegraphics[width=0.45\textwidth,clip]{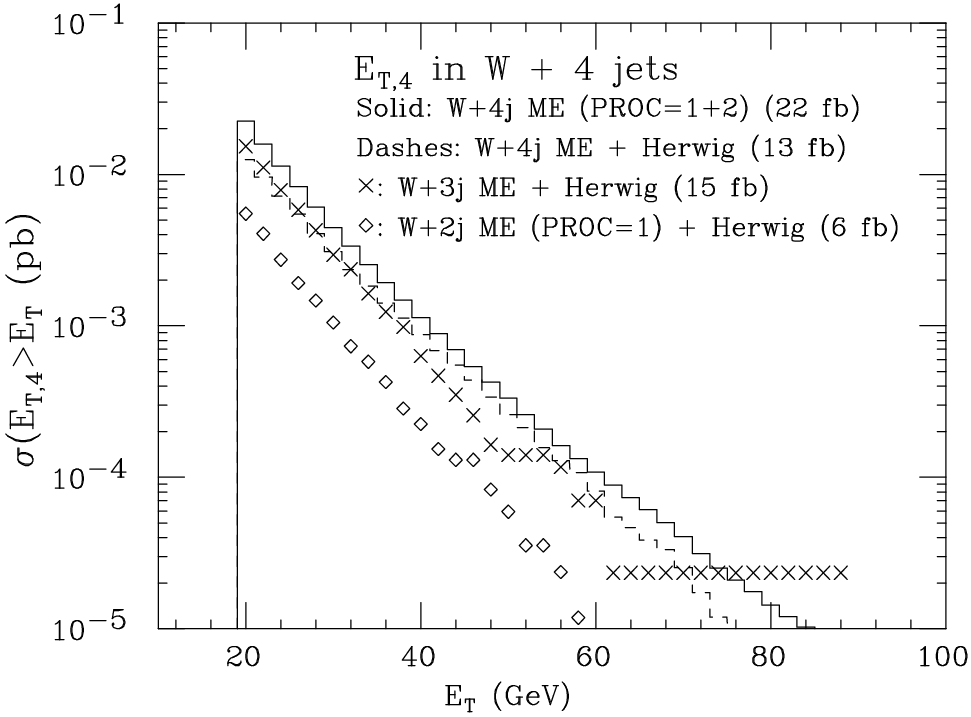}
\\
\includegraphics[width=0.45\textwidth,clip]{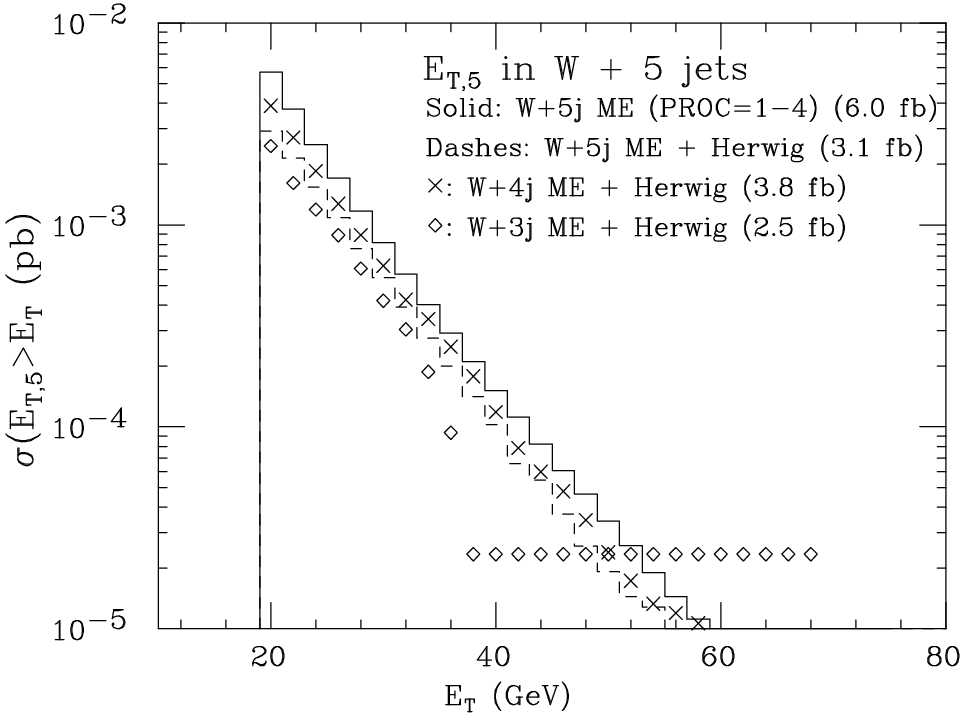}
\hfill
\includegraphics[width=0.45\textwidth,clip]{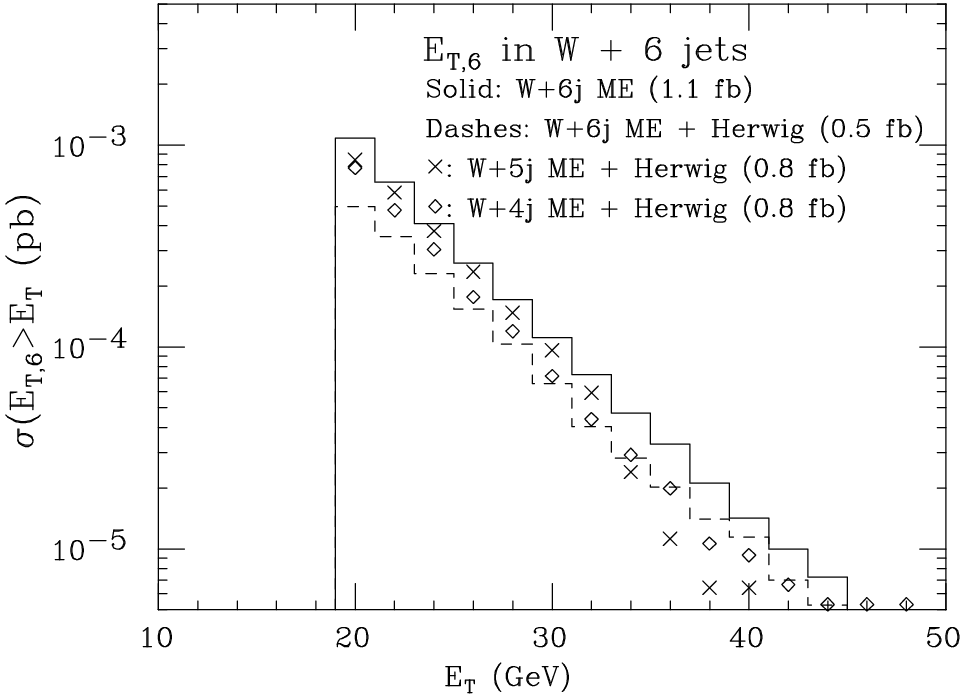}
\ccaption{}{\label{fig:2jhwpt3} Comparison of jet rates evaluated with
  matrix elements with those obtained from hard radiation during the
  shower evolution of lower-order parton-level processes.}
\end{center}
\end{figure}

We address in this section the issue of the ability of the shower MC to
correctly predict the rate for hard radiation leading to extra
final-state jets. The plots in fig.~\ref{fig:2jhwpt3} show the
spectrum of the 
$(N_J+1)$-th jet in events obtained by evolving through \herwig\
parton-level events with $N_J$ jets. These spectra are compared with
what obtained by using directly the $(N_J+1)$-jet parton-level matrix
elements, before (solid lines) and after (dashed lines) the \herwig\
evolution. As we pointed out in a previous section, large
contributions arise when higher-order  parton level processes are
considered. These contributions are due to the appearance of new processes,
which cannot be generated via radiation processes in the MC (e.g. new
initial states not present at the lower orders). In order to make a
fair comparison, we therefore included in the parton-level estimate of
the $(N_J+1)$-th jet spectrum only those processes which are present at
order $\as^{N_J}$. In the case of 3 final-state jets, we find a good
agreement in the range $\pt\lsim 45$~GeV 
between the rates obtained by evolving 3-jet parton level
events through \herwig\ (plotted points) 
and the rates obtained via radiation during the
shower evolution of 2-jet parton-level events (dashed line). 
A similar result is obtained in the case of 4 jets. In this case we
also studied the prediction of \herwig\ based on 2-jet parton-level
events, probing therefore the ability to predict the rate for the
emission of 2 extra jets (diamond symbols). The deficit in 4 jets
predicted using the 2-jet matrix elements is in part due to the lack of
$qg$-initiated processes, which  account for 35\% of
the 3-jet rate. Even including this correction, however, we see that
not enough hard radiation is emitted to correctly predict the emission
of 2 extra jets. The situation is improved when considering 5 and 6 jets, as
shown in the last frames of fig.~\ref{fig:2jhwpt3},
presumably because the overall amount of energy present in the events
in this case improves the validity of the soft-gluon emission
approximation even in the case of radiation leading to emission of 2
extra jets.

\section{Conclusions and outlook}
In this work we presented a 
general framework  for the 
 evaluation of complex multiparton matrix elements, including an
 algorithm for the consistent 
 merging with a coherent shower evolution, leading to fully
    hadronized multijet final states. We applied these ideas to the
 specific case of 
  $Wb\bar{b}+n$~jet production, completing the calculation for
 processes with up to a total of 6 final-state partons (including the $b$
 and $\bar{b}$). We
 presented results both at the parton level, and after the shower
 evolution, which was done using the \herwig\ MC. In addition to providing a
 proof of feasibility for these calculations and to showing the
 consistency of the merging with the coherent shower evolution, we
 pointed out several important consequences of the higher-order
 calculations and of the ability to maintain a non-zero mass for the
 $b$ quarks. On one side we stressed the fact that jet rates do not
 obey the naive $\as$ scaling with multiplicity, due to the appearance
 at higher orders of new, large contributions. This is especially true in
 the case of the lowest-order $W b\bar{b}$ and $W b\bar{b}+j$
 processes, and it is even more remarkable in the case of $pp$ collisions
 at the LHC.
 On the other side, we pointed out that measurements done by requiring
 the presence of 
 only one tagged \bjet\ are very sensitive to the contribution of
 \onebj\ events, namely events where the \bjet\ contains both the
 $b$ and the $\bar{b}$ quarks, or where one of the two $b$
does not reconstruct a
 taggable jet. The impact of these events on the background
 subtraction needed to determine the top cross section from the
 current measurements performed at the Tevatron remains to be
 assessed.

The formalism we have outlined, and the key use of \ALPHA\ to enable
the calculation of the matrix elements and the extraction of the colour
information needed for the coherent shower evolution, are readily
portable to the study of other complex multiparton processes. The
removal of the $b\bar{b}$ pair from the calculation will immediately
lead to the evaluation of $W+n$~jet rates, with $n$ up to 4. While
these processes have already been studied in the literature, and have
been encoded in the \vecbos\ MC, our approach will add to the existing
tools the ability to consistently evolve the final states through the
shower MCs, and to generate events suitable for realistic detector
simulations. The work on the $W+n$~jet program is in progress. If one
can neglect contributions from subprocesses with 6  or more light
quarks, an extension to $n$ up to 6 will be readily available using the
\ALPHA\ code. Inclusion of the processes with 6 or more light quarks
presents no conceptual problems, but simply requires a more involved
bookkeeping of all possible flavour channels.

The calculation of the 
processes $\ppbar \to t \bar{t} b \bar{b} + n$~jets and 
$\ppbar \to b \bar{b} b \bar{b} + n$~jets (with $n$ up to 2) is being
completed~\footnote{M.L. Mangano, G.~Montagna, M.~Moretti,
  O.~Nicrosini, F.~Piccinini and R.~Pittau, work in progress.}, and
will soon be reported on. Future work will then include the study of 
 $Z b\bar{b}+n$~jets and of $t\bar{t}+n$~jets.

\section*{Acknowledgements}
MM and RP thank the CERN Theory Division for hospitality during
various stages of this work.

\end{document}